\documentclass[12pt]{article}
\usepackage[utf8]{inputenc}
\usepackage{comment}
\usepackage{cite}
\usepackage[absolute]{textpos}
\input epsf.sty
\usepackage{cite}

\usepackage[margin=2cm]{geometry}
\usepackage{verbatim}         	
\usepackage{graphicx}				
\usepackage{cite}	
\usepackage{hyperref}
\usepackage{amssymb}
\usepackage{amsmath}
\usepackage{xcolor}
\newcommand{\be}{\begin{eqnarray}\displaystyle}
\newcommand{\ee}{\end{eqnarray}}
\newcommand{\nn}{\nonumber}

\newcommand{\p}{\partial}

\hypersetup{
  colorlinks   = true, 
  urlcolor     = blue, 
  linkcolor    = blue, 
  citecolor   = red  
}

\title{New Asymptotic Conservation laws for Electromagnetism.}
\author{}\date{}
\begin{document}	
\begin{textblock}{5}(6,1)
 \color{red}\Large $||$ Sri Sainath $||$
\end{textblock}
\color{black}
\maketitle
\centerline{\large {  Sayali Atul Bhatkar,  }}

\vspace*{4.0ex}

\centerline{\large \it Indian Institute of Science Education and Research,}
\centerline{\large \it  Homi Bhabha Rd, Pashan, Pune 411 008, India.}

\vspace*{1.0ex}
\centerline{\small E-mail: sayali014@gmail.com.}
\vspace{3cm}
\textbf{Abstract}\\

We obtain the subleading tail to the memory term in the late time electromagnetic radiative field generated due to a generic scattering of charged bodies. We show that there exists a new asymptotic conservation law which is related to the subleading tail term. The corresponding charge is made of a mode of the asymptotic electromagnetic field that appears at $\mathcal{O}(e^5)$ and we expect that it is uncorrected at higher orders. This hints that the subleading tail arises from classical limit of a 2-loop soft photon theorem. Building on the $m=1$ \cite{1903.09133, 1912.10229} and $m=2$ cases, we propose that there exists a conservation law for every $m$ such that the respective charge involves an $\mathcal{O}(e^{2m+1})$ mode and is conserved exactly. This would imply a hierarchy of an infinite number of $m$-loop soft theorems. We also predict the structure of $m^{th}$ order tails to the memory term that are tied to the classical limit of these soft theorems. 
\newpage

\tableofcontents
\vspace{2cm}
\section{Introduction}

Soft theorems are universal statements about quantum amplitudes in the limit when energy of one of the scattering particles is taken to be small \cite{soft0,soft1,soft2,soft3,soft4,soft5}. Soft theorems are related to asymptotic symmetries \cite{fer1,fer2,qed1,qed2,qed3,g1,g2,g3,g4,g5,sub1,sub2,sub3}.  Strominger and his collaborators provided a new insight into this picture. They proposed an equivalence between soft theorems and asymptotic conservation laws. In the classical theory, these conservation laws take following form : 
\be
Q^+[\lambda^+]\ |\ _{\mathcal{I}^+_-}\ \  = \ \ Q^-[\lambda^-]\ |\ _{\mathcal{I}^-_+}.\label{cons}
\ee
$\mathcal{I}^+$ is the future null infinity and $u=t-r$ is its null generator. The future charge $Q^+$ is defined at $\mathcal{I}^+_-$ i.e. the $u\rightarrow-\infty$ sphere of $\mathcal{I}^+$.  Similarly, $\mathcal{I}^-$ is the past null infinity and $v=t+r$ is the null generator. The past charge $Q^-$ is defined at $\mathcal{I}^-_+$ which is the $v\rightarrow\infty$ sphere of $\mathcal{I}^-$. $\lambda^+(\hat{x})$ parameterises the transformation and is an arbitrary function on $S^2$. The parameter at $\mathcal{I}^-_+$ is related to it via antipodal map $\lambda^+(\hat{x})=\lambda^-(-\hat{x})$. In \cite{qed1,qed2}, the authors discussed the asymptotic conservation law corresponding to a subgroup of the U(1) gauge group called large gauge transformations. They showed that the corresponding Ward identity for $S$-matrix : $Q^+S-SQ^-=0$ is equivalent to the leading soft theorem. This asymptotic conservation law was proved in \cite{lead asym}. The authors evolved the charge from $\mathcal{I}^-_+$ to $\mathcal{I}^+_-$ using classical equations of motion at spatial infinity and showed that it is indeed conserved.

It has been established that tree level amplitudes in QED admit subleading soft theorems.  An infinite number of tree level soft theorems have been proved in \cite{Hamada Shiu, soft inf}\footnote{The subleading terms admit corrections in presence of non-minimal couplings \cite{non uni}.}. In \cite{infinite asym}, the authors proved an infinite number of conservation laws for classical electromagnetism and also provided evidence that suggests that these conservation laws are equivalent to the soft theorems given in \cite{Hamada Shiu, soft inf}. Thus, tree level soft theorems in QED can be related to asymptotic conservation laws. While the leading soft photon theorem does not receive any loop corrections, \cite{loop1,loop2,loop3} showed that there are loop corrections beyond the leading order term in four spacetime dimensions. In \cite{Sen Sahoo}, the authors derived the subleading soft theorem for loop amplitudes and showed that it is 1-loop exact. The subleading term is logarithmic in soft energy and is absent in the tree level analysis. It is intimately tied to the long range forces present in four spacetime dimensions. In this paper, our aim is to study asymptotic conservation laws related to such kind of loop corrections.

An interesting aspect is that the leading soft theorem controls the leading order late time radiation emitted in any classical scattering process. The leading order term at late times is $\mathcal{O}(u^0)$ where $u$ is the retarded time. This term has been studied extensively - it is the so called memory effect \cite{mem1,mem2,mem3,mem4, Garfinkle}. The relation between the memory term and the leading soft theorem was established in \cite{Grav disp memory, Pasterski mem}. This relation between soft theorems and classical radiation was futher extended in \cite{G waves}. In \cite{G waves}, the authors proposed a novel way of taking classical limit of quantum multiple soft theorems \cite{multiple} and showed that soft theorems control the low energy (soft) radiation emitted in classical processes. 

It is natural to ask if the $\log \omega$ soft theorem is related to the late time radiation emitted in classical processes. It is expected that the analysis of \cite{G waves} can be repeated in this context to derive the classical limit of this soft theorem. In \cite{log waves, log mem}, the authors took another approach and directly studied the classical radiation emitted in some special setups. They showed that including the effect of long range forces leads to a new subleading term in the late time radiation :
\begin{align}
A_\mu|_{\mathcal{I}^+}=\frac{1}{4\pi r} \Big[\ a_\mu^\pm u^0 + \frac{b_\mu^\pm}{u} +...\ \Big],\ \ \  u\rightarrow\pm\infty.
\end{align}
The $u^0$-term is the memory term. It is controlled by the leading soft factor. The $\frac{1}{u}$-term is the leading tail to the memory term. It was calculated in \cite{log waves} and shown to be related to the $\log \omega$ soft factor. This term being a direct consequence of the long range electromagnetic force appears at $\mathcal{O}(e^3)$. In \cite{log mem em}, the authors extended the calculations to a general scattering process and showed that the results of \cite{log waves} hold in general i.e. $b_\mu^\pm$ are univeral. It is expected that this term is uncorrected by higher order corrections in $e$. The authors of \cite{log mem em} have also given a prediction for the expression of the subleading tail ($\frac{\log u}{u^2}$) to the gravitational memory term.

In this paper, we first study the subleading tail to the electromagnetic memory term. The subleading tail appears at $\mathcal{O}(e^5)$. We compute the explicit form of this $\frac{\log u}{u^2}$-term in the late time electromagnetic radiation emitted in a generic classical scattering of charged bodies. The structure of this term also turns out to be universal. This term is completely fixed by long range interaction (i.e. the $\frac{1}{r^2}$ term in asymptotic electromagnetic field) and depends only on the asymptotic momenta and charges of the scattered particles. It is insensitive to the details of scattering and is unchanged in presence of non-minimal couplings as well. We expect that this term is uncorrected  even when we go to higher orders in $e$. The universal nature of this term hints that this term should be tied to a soft theorem. 

To summarise, the first three terms in late time expansion of electromagnetic radiation are universal. The first one i.e. the memory term is related to an asymptotic conservation law\cite{qed1,qed2}. In \cite{1903.09133, 1912.10229}, the authors have proposed an asymptotic conservation law such that the charge is related to a logarithmic mode that appears at $\mathcal{O}(e^3)$. The Ward identity is equivalent to the 1-loop exact $\log\omega$ soft theorem and in the classical theory this charge controls the leading tail term. On the same lines we ask the question : is the subleading tail related to a new conservation law?

 In this paper, we start by proving above $\mathcal{O}(e^3)$ asymptotic conservation law for a general scattering process. 
More interestingly, we demonstrate existence of a new asymptotic conservation law such that charge is related to modes \footnote{See \eqref{2loop} for the precise form of the charge} of the asymptotic electromagnetic field that appear at $\mathcal{O}(e^5)$. This conservation law is related to the $\frac{\log u}{u^2}$-term in the late time electromagnetic radiation and we expect that in the quantum theory it would lead to a new soft photon theorem at 2-loop order. Based on the $m=1,2$ cases, we go ahead and propose an asymptotic conservation law for every $m$; the corresponding charge made of $\mathcal{O}(e^{2m+1})$ mode. The detailed form of these conservation laws is given in \eqref{ncons}. We expect that these charges do not get corrected by higher orders in $e$ and that these conservation laws should hold exactly in the full non-perturbative theory. We leave it to further investigations to prove these $\mathcal{O}(e^{2m+1})$ conservation laws following the analysis of \cite{infinite asym} in presence of long range forces. Existence of these infinite number of conserved asymptotic charges $Q_m$ at $\mathcal{O}(e^{2m+1})$ hints towards existence of a hierarchy of an infinite number of soft theorems for QED. Ward identity of the charge $Q_m$ would correspond to a new soft theorem at every $m$-loop order that is $m$-loop exact.

Let us briefly discuss the implications of our proposal at the level of classical theory. We expect that there exist following universal class of tails in the late time radiation :
\begin{align}
A_\mu=\frac{1}{4\pi r} \Big[\ a_\mu^\pm u^0\  +\ \sum_{\substack{m=1}}^\infty [b^{(m-1)}_\mu]^\pm \frac{(\log u)^{m-1}}{u^m}\ +...\Big] ,\ \ \  u\rightarrow\pm\infty.
\end{align}
We predict that the form of coefficients is given by :
\begin{align}
[b^{(m-1)}_\mu]^+ =\sum_{i=n'+1}^n \Big[\ Q_i  \ [\frac{p_{i\mu}}{ q.p_i} (q.c_i)-c_{i\mu} ]\ (q.c_i)^{m-1} + q_{\nu_1}\cdots q_{\nu_{m-1}}\ \mathcal{F}_{i\mu}^{\nu_1\cdots \nu_{m-1}}\ \Big],\nn\\
[b^{(m-1)}_\mu]^- =\sum_{i=1}^{n'} \Big[\ Q_i  \ [\frac{p_{i\mu}}{ q.p_i} (q.c_i)-c_{i\mu} ]\ (q.c_i)^{m-1} + q_{\nu_1}\cdots q_{\nu_{m-1}}\ \mathcal{F}_{i\mu}^{\nu_1\cdots \nu_{m-1}}\ \Big].\label{sn}
\end{align}
$i=1$ to $n'$ label the incoming particles and $i=n'+1$ to $n$ label the outgoing particles. $Q_i, p_i$ are respectively the asymptotic charges and momenta of the scattering bodies. $Q_i, p_i$ are defined including $\eta_i$ factors such that $\eta_i=1 (-1)$ for outgoing (incoming) particles. $c_i$'s given in \eqref{c} represent the effect of the long range force as seen in eq.\eqref{x1}. The form of $\mathcal{F}$ is fixed by long range force and the schematic form of this term is given in \eqref{FF}. We do not have explicit form of this term at every order but it can obtained in a straightforward (but tedious) way order by order. We have explicitly obtained the form of $b_\mu^{(m-1)}$ for $m=2$ in this paper. $m=1$ was already studied in \cite{log mem, log mem em}. We have also verified our proposals for $m=3$ and we plan to discuss it elsewhere.

Let us discuss the outline of this paper. In section 2, we review some known results to set up the background : we discuss the memory term, its relation to the leading soft factor and the corresponding conservation law. In section 3, we include the effects of long range electromagnetic force and discuss the resultant modes in the asymptotic field. Then we prove the $m=1$ conservation law. In section 4, we go to higher order in $e$ and obtain the subleading tail to the memory term.  We discuss some interesting aspects of this term. We show existence of $m=2$ conservation law in section 5. In section 6, we state our proposal for general $m$ and then summarise our results in section 7.

\section{Preliminaries}

Our aim is to study late time radiation emitted in a general classical scattering process. In a general scattering problem we have some $n'$ number of charged bodies coming in to interact. Let us denote the respective velocities by $V_i^\mu$, charges by $e_i$ and masses by $m_i$ (for $ i=1\cdots n'$). The interaction includes other kind of short range forces that could be present between charged bodies in addition to the electromagnetic force. $(n-n')$ number of final charged bodies with velocities $V_i^\mu$, charges $e_i$ and masses $m_i$ (for $ i=n'+1\cdots (n-n')$) repectively are produced as a result of the interaction. We assume that the non-electromagnetic forces die off faster than any power law in the asymptotic region. The only interaction between the particles in the asymptotic region is the electromagnetic interaction and starts at $\mathcal{O}(\frac{1}{r^2})$. Depending on the strength of the short range interactions, we can always choose a sphere of radius $T$ around the origin $r=0$ such that the short range forces can be ignored for $r>T$. We refer to this region as the asymptotic region. 

We need to calculate the asymptotic radiative field generated in such a generic process. We will carry out the calculations perturbatively in coupling $e$ as well as in asymptotic parameters $1/r$ (or $1/t$). It turns out that the internal structure of the scattering bodies is not relevant for our calculations. The current corresponding to an extended object can be written as the contribution from pointlike object plus corrective terms that depend on internal structure of the object like its charge distribution. These correspond to higher order moments and are subleading at large $r$. These corrections originating from internal structure of the scattering objects do not contribute to the modes that are of interest to us. We illustrate this in Appendix \ref{int} by showing that dipole term does not affect the subleading tail term in \eqref{Alogu} and the $Q_1, Q_2$ charges given in \eqref{1loop} and \eqref{2loop} respectively. Similarly we show in Appendix \ref{int} that these modes are not affected by non-minimal couplings.\footnote{ The coupling of internal spin to the gauge field is a special type of  non-minimal coupling. Hence our analysis is unaffected by internal spin of scattering objects. See \ref{int} for details.} Hence, we can study scattering of minimally coupled, spin zero point particles without any loss of generality. 
\color{black}
 
\subsection{The memory term}
In this section we will obtain the electromagnetic memory term \cite{mem4,Garfinkle,Pasterski mem} in the radiative field. We consider a test charge placed at a distance which is sufficiently far from the 'scattering region' which of size '$T$'. This test source will act as our detector for radiation. 
Radiation reaches $\mathcal{I}^+$  in the far furture. $\mathcal{I}^+$ corresponds to the limits $r\rightarrow\infty$ with $t-r$ finite. We use retarded co-ordinate system to describe  $\mathcal{I}^+$. The flat metric takes following form in this co-ordinate system ($u=t-r$) :
\be 
ds^2 = -du^2 - 2dudr + r^2\ 2\gamma_{z\bar{z}}\ dz d\bar{z}; \ \ \gamma_{z\bar{z}} = \frac{2}{(1+z\bar{z})^2}.\nn
\ee
We use $\hat{x}$ or $(z,\bar{z})$ interchangeably to describe points on $S^2$. We will often use following parametrisation of a 4 dimensional spacetime point :
\be x^\mu = rq^\mu + u t^\mu,\ \ \  q^\mu=(1,\hat{x}), \ \ \  t^\mu=(1,\vec{0}).\label{q}\ee
Here, $q^\mu$ is a null vector that can be parameterised in terms of $(z,\bar{z})$ as $$q=\frac{1}{1+z\bar{z}}\lbrace1+z\bar{z},z+\bar{z},-i(z-\bar{z}),1-z\bar{z}\rbrace.$$
In Lorenz gauge, the radiation can be obtained from the equation $\Box A_\mu=-j_\mu$. Using the retarded propagator, we get :
\begin{align}
A_\sigma(x)
&=\frac{1}{2\pi }\int d^4x'\ \delta([x-x']^2)\ j_\sigma(x')\  \Theta(t-t') .
\end{align} 
Let us study the retarded root of the delta function $\delta([x-x']^2)$ which is given by 
\begin{align}
t'_0&=t-|r-r'|\nn.
\end{align}
The form of $t'_0$ at large $r$ is $t'_0=u+\mathcal{O}(\frac{1}{r})$. Thus the field $A_\sigma(r,u,\hat{x})$ at large $r$ gets contribution from $t'\sim u$. The bulk region corresponds to $r'<T$ i.e. $|t'|<T$ as we have set $c=1$. Hence it contributes to $A_\sigma$ at $|u|<T$. The asymptotic field at large $u$ does not get contribution from the bulk region $|t'|<T$. Thus to find $A_\sigma$ near $\mathcal{I}^+_\pm$, we can focus only on the asymptotic ($t'>T$) trajectories. Let us first restrict ourselves to the leading order in coupling $e$, then we can ignore the effect of long range electromagetic interactions on the asymptotic trajectories. Hence the particles are free asymptotically (i.e. for $r>T$). Thus an incoming particle has the trajectory (Greek indices will be used to denote 4d cartesian components) :
$$ x^\mu_i= [V_i^\mu \tau + d_i]\Theta(-T-\tau).$$
$\tau$ is an affine parameter, here $T$ denotes the value of $\tau$ such that the short range forces can be ignored for $\tau>T$. Similarly, an outgoing particle has the trajectory :
$$ x^\mu_j= [V_j^\mu \tau + d_j]\Theta(\tau-T).$$
The bulk trajectories might have any complicated form depending on the short range forces, that will not affect our analysis. In the asymptotic region, the trajectories are simpler and given by above expressions (upto corrections in $e$). The current is given by summing over all particles that participate in the scattering. The asymptotic part of this current can be written down as :
$$j^{\text{asym}}_\sigma(x') = \int d\tau\Big[\sum_{i=n'+1}^n e_i V_{i\sigma}\  \delta^4(x'-x_i)\ \Theta(\tau-T)+\sum^{n'}_{i= 1} e_i V_{i\sigma}\  \delta^4(x'-x_i)\ \Theta(-T-\tau)\Big].$$
Here, we have labelled the incoming particles by $i$ running from 1 to $n'$ and outgoing particles by $i$ running from $n'+1$ to $n$. Next we need to find the radiation produced by above current. Using the retarded propagator, we get :
\begin{align}
A^{\text{asym}}_\sigma(x)
&=\frac{1}{2\pi }\int d^4x'\ \delta([x-x']^2)\ j^{\text{asym}}_\sigma(x')\  \Theta(t-t') .\label{A1}
\end{align}
We have added a superscript to note that we have ignored the bulk sources of radiation. Henceforth, we will drop this superscript but it should be always remembered that we are calculating only the asymptotic part of the field. To avoid clutter, let us first consider the electromagnetic field generated by the $i^{th}$ asymptotically free outgoing particle. At the end, we just need to sum over all particles. We denote it with a superscript $(i)$.
\begin{align}
A^{(i)}_\sigma(x)
&=\frac{1}{2\pi }\int d^4x'\ \delta([x-x']^2)\ j^{(i)}_\sigma(x')\ \Theta(t-t')\ , \nn\\
&=\frac{1}{2\pi }\int d\tau\ \frac{\delta(\tau-\tau_0)}{|2\tau+2V_i.(x-d_i)|}\  e_iV_{i\sigma}  \Theta(\tau-T).
\end{align}
The retarded root of the delta function $\delta([x-x']^2)$ is given by 
\be \tau_0=-(V_i.x-V_i.d_i)-\sqrt{(V_i.x-V_i.d_i)^2+(x-d_i)^2}.\label{tau0}\ee
Hence,
\begin{align}
A^{(i)}_\sigma(x)
&=\frac{1}{4\pi }  \frac{e_iV_{i\sigma}\ \Theta(\tau_0-T) }{\sqrt{(V_i.x-V_i.d_i)^2+(x-d_i)^2}}  .\nn
\end{align}
Thus, we can write down the total asymptotic field generated by the scattering process. It is given by :
\begin{align}
A_\sigma(x)
&=\sum_{i=n'+1}^n\frac{1}{4\pi }  \frac{e_iV_{i\sigma}\ \Theta(\tau_0-T) }{\sqrt{(V_i.x-V_i.d_i)^2+(x-d_i)^2}}  +\sum_{i=1}^{n'}\frac{1}{4\pi }  \frac{e_iV_{i\sigma}\ \Theta(-T-\tau_0) }{\sqrt{(V_i.x-V_i.d_i)^2+(x-d_i)^2}}\label{A0} \ .
\end{align}

To find the radiative field around $\mathcal{I}^+$ let us take the limit $r\rightarrow\infty$ with $u$ finite in eq.\eqref{A0}. Using $\tau_0= \frac{u}{|q.V_i|}+\mathcal{O}(1)$ in \eqref{A0} :
\begin{align}
A_\sigma(x)|_{\mathcal{I}^+}
&=-\frac{1}{4\pi r}\Big[\sum_{i=n'+1}^n  \frac{e_iV_{i\sigma}}{V_i.q} \Theta(u-T) +\sum_{i=1}^{n'}\frac{1}{4\pi }  \frac{e_iV_{i\sigma}}{V_i.q}  \Theta(-T-u) + ...\Big]+\mathcal{O}(\frac{1}{r^2})\ . \label{A00}
\end{align}
The $\frac{1}{r}$-term gives us the radiative field. At large values of $u$, we see that it goes like $u^0$. '$...$' denote $u$-fall offs that are faster than any (negative) power law behaviour. Let us rewrite \eqref{A00} a bit succinitly :
\begin{align}
A_\mu(x)=\frac{1}{4\pi r} \ a_\mu^\pm(\hat{x}) u^0\  +...,\ \ \  u\rightarrow\pm\infty.\label{AA1}
\end{align}
Let us discuss the effect of this mode on the test particle. Our test particle is placed at large $r$ and large $u$. Let us denote its  trajectory by $x^\mu(\sigma)$. The equation of motion is given by :
\begin{align}
m\frac{\p V^\mu}{\p \sigma}=eF^\mu_{\nu}V^\nu.
\end{align} 
 Here, $V^\mu = \frac{\p x^\mu}{\p \sigma}$. We will use capital latin alphabets to denote indices on $S^2$. These indices take values $(z,\bar{z})$. Using a co-ordinate transformation we have : $V_A= (\p_A x^\mu)V_\mu$. We use \eqref{A00} to find the leading order term in $F_{\mu\nu}V^\nu$.
We get (noting that $F^{[r^0]}_{AB}=0$) :
\begin{align}
m\frac{\p V_A}{\p \sigma}=e\ \p_uA_A^0 V^u+\mathcal{O}(\frac{1}{r}).
\end{align} 
In above equation, $A_A^0$ is used to denote following mode in $A_A(x)$ : $A_A(x)\sim A_A^0(u,\hat{x}) + \mathcal{O}(\frac{1}{r})$. We can easily solve above equation to get the net change in velocity. This velocity shift of the test particle is given by : $\Delta V_A = \frac{e}{m} \Delta A_A^0.$
We note that this shift is in a plane transverse to $\hat{r}$. Using \eqref{A00}, we can calculate the shift in gauge field explicitly.
\be \Delta A_A^0(\hat{x})\ \ =\ \ \int_{\mathcal{I}^+} du\ \p_u A_A^0(u,\hat{x})\ \ =\ \  -\frac{1}{4\pi}\Big[ \sum_{i=n'+1}^n e_i\frac{ (\p_A q).V_i}{V_i.q}-\sum_{i=1}^{n'} e_i\frac{ (\p_A q).V_i}{V_i.q}\Big], \label{deltaA} \ee
here, we have labelled the incoming bodies by $i$ running from 1 to $n'$ and outgoing bodies by $i$ running from $n'+1$ to $n$. We note that $\Delta V_A$ has a nice form and it is insensitive to the details of scattering. 
Let us use the oft-used basis for polarisation vectors \cite{qed1}:
\be \epsilon^\mu_- = \frac{1}{\sqrt{2}}\frac{\p}{\p \bar{z}} [(1+z\bar{z})q^\mu],\ \  \epsilon^\mu_+ = \frac{1}{\sqrt{2}}\frac{\p}{\p z} [(1+z\bar{z})q^\mu].\label{pol}\ee
Then we can write \be \Delta A_z^0(\hat{x})\ \ =\ \  -\frac{1}{4\pi} \sum_{i=1}^n Q_i\frac{ \epsilon_+ .p_i}{p_i.q}, \ee 
Here, we have used the convention that is commonly used for asymptotic quantities : 
\be Q_i=\eta_ie_i\text{ and } p_i=\eta_im_iV_i, \label{eta}\ee
such that $\eta_i=1 (-1)$ for outgoing (incoming) particles. The expression for $A_{\bar{z}}$ is given by replacing $\epsilon^\mu_-$ by $\epsilon^\mu_+$.

To summarise we obtained the asymptotic field generated by a scattering event at zeroth order ignoring the long range forces acting on the scattering particles. In particular we studied the effect of the leading order $u^0$-term on a test particle (placed at large $r$ and large $u$). The velocity of the test particle receives a kick on the passage of electromagnetic radiation due to the $u^0$-term. This is the so called memory effect. As visible from the form of the expression, the amount of kick received is insensitive to the details of scattering and in fact is proportional to the leading soft factor. This has been amply discussed in the literature \cite{Garfinkle, Pasterski mem}. Let us comment on corrections from terms higher order in $e$. The $u^0$ does not get any corrections from higher order terms in $e$.  In \eqref{AA1}, we have :
\be A_\mu(x) \sim \ \frac{1}{4\pi r}\Big[ a_\mu^\pm(\hat{x}) u^0 + ...\Big] +\mathcal{O}(\frac{1}{r^2}). \label{falloff}\ee
We will see in the forthcoming sections that $a_\mu^\pm$ is uncorrected even when we go to higher orders in $e$. The subleading behaviour is changed substantially as we go to next order in $e$.

\subsection{Conservation law for the leading charge $Q_0$}
Next we will show that particular mode in the field strength satisfies an asymptotic conservation law. This law is related to the leading soft theorem. We recall that asymptotic charges are conserved in following sense :
\be
Q^+[\lambda^+]\ |\ _{\mathcal{I}^+_-}\ \  = \ \ Q^-[\lambda^-]\ |\ _{\mathcal{I}^-_+}.\label{Qcon}
\ee
The future charge $Q^+$ is defined at $\mathcal{I}^+_-$ i.e. the $u\rightarrow-\infty$ sphere of $\mathcal{I}^+$. Similarly, the past charge $Q^-$ is defined at $\mathcal{I}^-_+$ which is the $v\rightarrow\infty$ sphere of $\mathcal{I}^-$. Let us calculate the asymptotic field strength produced by the scattering event using \eqref{A00}. We get around $u\rightarrow -\infty$ :
\begin{align}
F_{\mu\nu}(x)|_{u\rightarrow -\infty}
&=\frac{1}{4\pi }\sum_{i=1}^{n'}   \frac{e_i \big[V_{i\mu}(x_{\nu}-d_{i\nu})-(x_{\mu}-d_{i\mu})V_{i\nu}\big]}{[(V_i.x-V_i.d_i)^2+(x-d_i)^2]^{3/2}}.  \label{F1}
\end{align}
It is important to recall that there will be corrections to above expression when we go to higher orders in $e$. These corrections would be subleading at large $u$. The leading order term in above expression is $\frac{1}{r^2}$. Using \eqref{q}, the coefficient of this term can be written down.
\begin{align}
F_{\mu\nu}(x)|_{u\rightarrow -\infty}
&=  -\frac{1}{4\pi r^2}\sum_{i=1}^{n'}   \frac{e_i  }{(V_i.q)^3}\big[V_{i\mu}q_{\nu}-q_{\mu}V_{i\nu}\big] +\mathcal{O}(\frac{1}{r^3}). 
\end{align}
Performing a co-ordinate transformation :
\begin{align}
F_{ur}(x)|_{u\rightarrow -\infty}
&=-\frac{1}{4\pi r^2}\sum_{i=1}^{n'}  \frac{e_i\ }{ (V_i.q)^2}  +\mathcal{O}(\frac{1}{r^3})\ \label{Fru} .
\end{align}
Next we need to derive the field configuration at past null infinity and then compare the two expressions. We need to take the limit $r\rightarrow\infty$ with $v=t+r$ finite. In this co-ordinate system, 4 dimensional spacetime point can be parametrised as :
\be x^\mu = r\bar{q}^\mu + v t^\mu,\ \ \  \bar{q}^\mu=(-1,\hat{x}), \ \ \  t^\mu=(1,\vec{0}).\label{qbar}\ee
Again, $\bar{q}^\mu$ is a null vector. We need to expand all the quantities around $\mathcal{I}^-$. Around $\mathcal{I}^-$, we have from \eqref{tau0} : $\tau_0= -2r\ V_i.\bar{q}+\mathcal{O}(1)$. Using \eqref{A0}, we see that outgoing particles do not contribute to the field at $\mathcal{I}^-$ because of $\Theta(\tau_0-T)$. This is a consequence of retarded boundary conditions.  From \eqref{A0}, we get :
\begin{align}
A_\sigma(x)|_{\mathcal{I}^-}
&=\frac{1}{4\pi r}\Big[\sum_{i=1}^{n'}  \frac{e_iV_{i\sigma}}{V_i.\bar{q}} + \frac{1}{v^\infty}\Big]+\mathcal{O}(\frac{1}{r^2})\ . 
\end{align}
Calculating the field strength,  we get 
\begin{align}
F_{vr}(x)|_{v\rightarrow \infty}
&=-\frac{1}{4\pi r^2}\sum_{i=1}^{n'}  \frac{e_i\ }{ (V_i.\bar{q})^2}  +\mathcal{O}(\frac{1}{r^3})\ .
\end{align}
Using \eqref{Fru} along with above equation, we have rederived the 'conservation' law \cite{qed1,lead asym} :
\begin{align}
F^{[u^0/r^2]}_{ur}(\hat{x})|_{\mathcal{I}^+_-}=
F^{[v^0/r^2]}_{vr}(-\hat{x})|_{\mathcal{I}^-_+}.
\end{align}
Here, $F^{[u^0/r^2]}_{ur}$ denotes the coefficient of $\frac{u^0}{r^2}$-term of $F_{ur}$. Hence it a function of the sphere co-ordinates $\hat{x}$. Similarly  $F^{[v^0/r^2]}_{ur}$ denotes the coefficient of $\frac{v^0}{r^2}$-term of $F_{vr}$. The future charge is defined as $Q^+_{0}[\lambda^+]=\int d^2z\ \lambda^+(\hat{x})\ {F^{[u^0/r^2]}_{ur}}(\hat{x})$. $Q_{0}^-$ is defined analogously. We have : 
\be
Q_0^+[\lambda^+]\ |\ _{\mathcal{I}^+_-}\ \  = \ \ Q_0^-[\lambda^-]\ |\ _{\mathcal{I}^-_+}.\label{Qcon}
\ee
 $\lambda^+$ is an arbitrary function on 2-sphere and $\lambda^+(\hat{x}) = \lambda^-(-\hat{x})$. The leading soft theorem can be understood as a Ward identity for $S$-matrix : $Q^+_{0}S-SQ^-_{0}=0$ \cite{qed1,qed2} with $Q_0$'s defined as given above.

\section{$m=1$ conservation law} 
In this section we will obtain the asymptotic radiative field keeping the first order correction in $e$. Therefore we need to take into account the effect of long range forces acting on the particles.  In four space-time dimensions these $\frac{1}{r^2}$-forces are subtle as they induce logarithmic correction to the straight line trajectory at late times. As a result of long range forces, a particle continues to accelerate at late times and this gives rise to new modes in the asymptotic field at $\mathcal{O}(e^3)$. 
In particular $F_{rA}$ gets a log $u$ term because of the long range interaction between particles : 
\be
F_{rA}|_{u \rightarrow -\infty} = \frac{1}{r^2}\ [\ u\ F^{[u/r^2]}_{rA}(\hat{x})\ +\ \log u\ F^{[\log u/r^2]}_{rA}(\hat{x})\ + ... ] +\mathcal{O}( \frac{1}{r^3})\ . \ee
Similarly around the past null infinity we have : 
\be
F_{rA}|_{v \rightarrow \infty} =  \frac{\log r}{r^2}\ [v^0 \ F^{[\log r/r^2]}_{rA}(\hat{x})\ +...]+ \mathcal{O}(\frac{1}{r^2})\ .\ee
These modes have been studied in \cite{1903.09133, 1912.10229} in the context of scalar fields. In this section we will derive the $\mathcal{O}(e^3)$ conservation law obeyed by these modes.
\begin{align}
F^{[\log u/r^2]}_{rA}(\hat{x})|_{\mathcal{I}^+_-}=
F^{[\log r/r^2]}_{rA}(-\hat{x})|_{\mathcal{I}^-_+}.
\end{align}
 This is the first of $\mathcal{O}(e^{2m+1})$ laws that we discuss. In this process we will also rederive the leading tail to the memory term.

Let us find the late time trajectory of a particle moving under the influence of long range forces that are exerted by other scattered particles. The leading order acceleration experienced by asymptotic particles was already calculated in \cite{Sen Sahoo} and has been rederived in Appendix \ref{lrf}. In \eqref{F1} we calculated the electromagnetic field in the asymptotic regime to $\mathcal{O}(e)$, using \eqref{F1} and \eqref{F} we see that the forces $F$ felt by a scattering particle takes following form : 
$$F(\tau) =e\ \sum_{m=2}^\infty\frac{c_m}{\tau^m}.$$
The $\mathcal{O}(\frac{1}{\tau^2})$-term solely depends upon the charge and asymptotic velocity. Dipole interactions and higher order moments contribute at $m>2$ and hence the $m>2$ modes are sensitive to the other details of the scattering objects. This has been discussed in Appendix \ref{int}. Due to the long range electromagnetic force, the equation of trajectory of an outgoing particle $i$ gets corrected to  :
\be x^\mu_i=V_i^\mu \ \tau +c^\mu_i \log \tau + d_i+\mathcal{O}(\frac{1}{\tau}). \label{x1} \ee
From \eqref{ca}, we have :
\begin{align}
c_i^\mu&=\frac{1}{4\pi} \sum_{\substack{j=m+1,\\ j\neq i}} ^n Q_iQ_j \frac{ p_i.p_j\ m_j^2 p^\mu_i+m_i^2m_j^2p_j^\mu}{[(p_i.p_j)^2-m_i^2m_j^2]^{3/2}}.\label{c}
\end{align}
An outgoing particles moves in the field of other outgoing particles hence the $c_i$'s for an outgoing particle includes contribution only from outgoing particles. Similarly an incoming particle can interact only with incoming particles via the long range forces. The effect of higher order moments of scattering objects enter \eqref{x1} at $\mathcal{O}(\frac{1}{\tau})$.

Next we find the asymptotic field produced by an outgoing particle $i$ with the corrected trajectory given in \eqref{x1}. As a result of long range interactions, the current corresponding to the particle is modified to $j^{(i)}_\sigma(x') = \int d\tau \ e_i \big[V_{i\sigma}+\frac{c_{i\sigma}}{\tau}\big]\  \delta^4(x'-x_i)\ \Theta(\tau-T).$ Using the retarded propgator we can write down the expression for radiation sourced by asymptotically accelerating particle.
\begin{align}
A^{(i)}_\sigma(x)
&=\frac{1}{2\pi }\int d\tau\ \delta([x-x'_i(\tau)]^2)\  e_i \big[V_{i\sigma}+\frac{c_{i\sigma}}{\tau}\big]\   \Theta(t-t') \  \Theta(\tau-T)
.\label{A1}\end{align}
This equation includes $\mathcal{O}(e^3)$ corrections to the eq.\eqref{A0}. Solving the $\delta$-function condition is highly difficult because of the logarithmic correction. We solve it perturbatively. The details of the calculation are relegated to Appendix A. We quote the solution to the delta function constraint from \eqref{tau1} :
\begin{align}
\tau_1&=-V_i.(x-d_i)-\big[\ (V_i.x-V_i.d_i)^2 +(x-d_i)^2\ -2(x-d_i).c_i\log\tau_0\ \big]^{1/2}.\label{t}
\end{align}
Here, $\tau_0$ is the zeroth order solution given in \eqref{tau0}. 
We also have $\delta([x-x'_i(\tau)]^2) = \frac{\delta(\tau-\tau_1)}{|2\tau+2V_i.(x-d_i)+\frac{2(x-d_i).c_i}{\tau} |}$. The result of the integral is :
\begin{align}
A^{(i)}_\sigma(x)
&=\frac{1}{4\pi }  \frac{\Theta(\tau_1-T)\ e_i \big[V_{i\sigma}+\frac{c_{i\sigma}}{\tau_1}\big] }{\big[\ (V_i.x-V_i.d_i)^2 +(x-d_i)^2\ -2(x-d_i).c_i\log\tau_0\big]^{1/2}-\frac{(x-d_i).c_i}{\tau_1}}.
\end{align}
Since above expression is vaild only to $\mathcal{O}(e^3)$, we can expand the denominator to $\mathcal{O}(e^3)$ as well. Summing over all the incoming and outgoing particles, we get :
\begin{align}
A_\sigma(x)
&=\frac{1}{4\pi } \sum_{i=n'+1}^n  \frac{e_i }{X}\Theta(\tau_1-T)\  \Big[\ V_{i\sigma} \big[1+\frac{1}{X^2}(x-d_i).c_i\log\tau_0+\frac{(x-d_i).c_i}{X\tau_1}\big]+\frac{c_{i\sigma}}{\tau_1}\Big]\nn\\
&+\frac{1}{4\pi } \sum_{i=1}^{n'}  \frac{e_i }{X}\Theta(-\tau_1-T)\  \Big[\ V_{i\sigma} \big[1+\frac{1}{X^2}(x-d_i).c_i\log\tau_0+\frac{(x-d_i).c_i}{X\tau_1}\big]+\frac{c_{i\sigma}}{\tau_1}\Big]\ ,\label{A01}
\end{align} 
\begin{align} 
&\text{where }X=[\ (V_i.x-V_i.d_i)^2 +(x-d_i)^2\ ]^{1/2}. \label{X}&&
\end{align} 
This is the main result at $\mathcal{O}(e^3)$. It is worth recalling that above expression is valid only in asymptotic regions, we have ignored the contribution of the bulk sources. 

Next we will expand above expression. Focussing on the $\frac{1}{r}$-term of $A_\sigma$, we get :
\begin{align}
A_\sigma(x)|_{\mathcal{I}^+}
&= -\frac{1}{4\pi r} \sum_{i=n'+1}^n Q_i  \Theta(u-T) \Big[\ \frac{p_{i\sigma}}{q.p_i} - \frac{1}{u}\big[c_{i\sigma}-p_{i\sigma}\frac{q.c_i}{q.p_i}\big] \Big]\nn\\
&+\ \frac{1}{4\pi r}\sum_{i=1}^{n'}Q_i  \Theta(-u-T) \Big[\  \frac{p_{i\sigma}}{q.p_i}-\frac{1}{u} \ \big[c_{i\sigma}-p_{i\sigma}\frac{q.c_i}{q.p_i}\big]\ \Big] +...\ .\label{1/u}
\end{align}
$p_i,Q_i$ have been defined in \eqref{eta}. Let us rewrite \eqref{1/u} a bit succinitly :
\begin{align}
A_\mu(x)|_{\mathcal{I}^+}=\frac{1}{4\pi r} \Big[\ a_\mu^\pm(\hat{x}) u^0\ +\ \frac{b_\mu^\pm(\hat{x})}{u}\ \Big] +...,\ \ \  u\rightarrow\pm\infty.\label{falloff1}
\end{align}
We can compare above fall offs to the leading order radiative fall offs in \eqref{falloff}. It is interesting to note that including even the first order correction in $e$ has altered the late time profile appreciably. This is the so called tail memory effect \cite{log mem, log mem em}. Let us note an important point about \eqref{falloff1}. We have ignored the subleading terms arising from the higher order moments in \eqref{x1}. If we include the effect of these subleading terms, then \eqref{falloff1} takes following form :
\begin{align}
A_\mu(x)\sim\frac{1}{4\pi r}\Big[ e\ u^0\ +\ e^3\sum_{n=1}^\infty\frac{1}{u^n}\ \Big]+\mathcal{O}(\frac{1}{r^2}) ,\ \ \  u\rightarrow\pm\infty.
\end{align}
In the next section, we will study $\mathcal{O}(e^5)$ corrections to above equation and show that there is a $\frac{\log u}{u^2}$ term which will dominate the $\mathcal{O}(\frac{1}{u^2} )$ terms. Another interesting aspect is that the coefficients of $u^0$ and $\frac{1}{u}$ are uncorrected by $\mathcal{O}(e^5)$ terms while other terms ($n >1$) get corrected at $\mathcal{O}(e^5)$.

Let us turn to the $\frac{1}{r^2}$-term of $A_\sigma$ and derive the conservation law we briefly discussed earlier. In the previous section, we rederived the conservation law that is equivalent to the leading soft photon theorem. We will follow similar strategy here.  We need to expand all the terms in \eqref{A01} around $\mathcal{I}^+$. Using \eqref{tau0}, we have $\log\tau_0|_{\mathcal{I}^+}\sim \log u+\mathcal{O}(u^0)$. Then using \eqref{q} and the fact that $V_i, d_i$ are $\mathcal{O}(r^0)$  parameters, we can find the leading order term in \eqref{t} :
\begin{align}
\tau_1|_{\mathcal{I}^+}&=-\frac{u}{q.V_i}- \ \frac{q.c_i}{q.V_i} \ \log [\frac{-u}{q.V_i}]+\mathcal{O}(1).
\end{align}
Using \eqref{q} in \eqref{X}, we get $X=-rq.V_i+\mathcal{O}(r^0)$.  Substituting the limiting value of $X$ in \eqref{A01}, we can read off the coefficient of the $\mathcal{O}(\frac{\log u}{r^2})$ term in $A_\sigma$ :
\begin{align}
A^{[\log u/r^2]}_\sigma(x)|_{\mathcal{I}^+}
&=-\frac{1}{4\pi } \sum_{i=n'+1}^n \Theta(u-T)\  e_i \ V_{i\sigma} \ \frac{q.c_i}{(q.V_i)^3}-\frac{1}{4\pi } \sum_{i=1}^{n'} \Theta(-u-T)  e_i \ V_{i\sigma} \ \frac{q.c_i}{(q.V_i)^3}.
\end{align}
From here on, we just need to transform co-ordinates to get to $F_{rA}$. We have : $A_r=q^\mu A_\mu$ and $A_A=r(\p_A q^\mu)A_\mu$. Using it in $F_{rA} = \p_rA_A-\p_A A_r$, we get :
\begin{align}
F^{[\log u/r^2]}_{rA}(x)|_{\mathcal{I}^+_-}
&=\frac{1}{4\pi }\sum_{i=1}^{n'}  \frac{e_i\ q^\mu (\p_A q^\nu ) }{(q.V_i)^3}  \ [V_{i\mu} c_{i\nu}-V_{i\nu} c_{i\mu}].\label{F1f}
\end{align}
Next we need to derive the field configuration at past null infinity and compare above expression with the coefficient of $\mathcal{O}(\frac{\log r}{r^2})$ term in $F_{rA}$. So, we expand $A_\sigma$ in \eqref{A01} around $\mathcal{I}^-$. Using \eqref{qbar} the leading order term in \eqref{t} is :
\begin{align}
\tau_1|_{\mathcal{I}^-}&=-2r\ V_i.\bar{q}+ \  \frac{\bar{q}.c_i}{V_i.\bar{q}} \ \log [-2r\ V_i.\bar{q}]+\mathcal{O}(1). \label{t1p}
\end{align}
Using \eqref{tau0}, we get $\log\tau_0|_{\mathcal{I}^-}\sim \log r+\mathcal{O}(r^0)$. Substituting in \eqref{A01}, we write down the coefficient of the $\mathcal{O}(\frac{\log r}{r^2})$ term in $A_\sigma$ :
\begin{align}
A^{[\log r/r^2]}_\sigma(x)|_{\mathcal{I}^-}
&=\frac{1}{4\pi } \sum_{i=1}^{n'} e_i \ V_{i\sigma} \ \frac{\bar{q}.c_i}{(\bar{q}.V_i)^3}.
\end{align}
Performing co-ordinate transformation :
\begin{align}
F^{[\log r/r^2]}_{rA}(x)|_{\mathcal{I}^-_+}
&=-\frac{1}{4\pi }\sum_{i=1}^{n'}  \frac{e_i\  \bar{q}^\mu (\p_A \bar{q}^\nu )}{(\bar{q}.V_i)^3}  \ [V_{i\mu} c_{i\nu}-V_{i\nu} c_{i\mu}].\label{F1p}
\end{align}
Thus, from \eqref{F1f} and \eqref{F1p} we can indeed check that the modes are equal under antipodal idenfication.  The apparently extra minus sign in \eqref{F1p} is compensated by the factors of $q^\mu$.
Finally we have shown that a generic scattering process obeys following conservation law :
\begin{align}
F^{[\log u/r^2]}_{rA}(\hat{x})|_{\mathcal{I}^+_-}=
F^{[\log r/r^2]}_{rA}(-\hat{x})|_{\mathcal{I}^-_+}.\label{1loop}
\end{align}
This is the $\mathcal{O}(e^3)$ conservation law. The charges are defined as $Q^+_{1}=\int d^2z \ F^{[\log u/r^2]}_{rA}(\hat{x})\ W^A(\hat{x})|_{\mathcal{I}^+_-}$ and $Q^-_{1}=\int d^2z\ F^{[\log r/r^2]}_{rA}(-\hat{x})W^A(-\hat{x})|_{\mathcal{I}^-_+}$. These charges are parametrised by an $S^2$ vector field $W^A$. 

Let us briefly discuss the implications of above conservation law on the quantum amplitudes. Loop amplitudes admit a subleading soft photon theorem :
\begin{align}
\mathcal{A}_{n+1}(p_i,k)&=\frac{{S}_0}{\omega}\ \mathcal{A}_n(p_i)\  + \ S_{\text{1-loop}}\log{\omega}\ \mathcal{A}_n(p_i)\ + ...\ .\label{soft1}
\end{align}
$S_{\text{1-loop}}$ has been derived in \cite{Sen Sahoo}. The $Q_1$ charges have been constructed for QED coupled to scalar fields in \cite{1903.09133, 1912.10229}. The authors showed that the corresponding Ward identity for $S$-matrix i.e. $Q^+_{1}S-SQ_{1}^-=0$ is equivalent to the 1-loop level $\log \omega$ soft theorem. In this paper, we showed that the charges constructed in these papers are indeed conserved in the classical theory.

It is interesting to compare the frequency space radiative field with \eqref{soft1}. Given \eqref{falloff1}, we can study its fourier transform. The Fourier transformed function has following behaviour at small $\omega$ \cite{log mem em} :
\be \tilde{A}_\mu(\omega,r,\hat{x}) =\frac{e^{i\omega r}}{4\pi ir}\Big[-\frac{1}{\omega}[a_\mu^+(\hat{x})-a_\mu^-(\hat{x})]-i[b_\mu^+(\hat{x})-b_\mu^-(\hat{x})]\ \log\omega +...\Big] \ \ \text{as }\omega\rightarrow 0.\label{class2} \ee
$\frac{e^{i\omega r}}{4\pi ir}$ is the overall normalization factor \cite{G waves}. The classical soft factor is :
\be S_1 =-i\epsilon^\mu[b_\mu^+-b_\mu^-]=i \sum_{j=1}^n Q_j\ \big[\frac{ \epsilon.p_j}{p_j.q}\ q.c_j-\epsilon.c_j\ \big]\ . \label{Slog}\ee
We note that the coefficient of $\log \omega$ in the classical field i.e. $S_1$ is only a part of $S_{\text{1-loop}}$ that appears in quantum soft theorem \eqref{soft1}. A part of $S_{\text{1-loop}}$ vanishes in the classical theory. This point was already discussed in \cite{Sen Sahoo}. It is expected that \eqref{class2} can be obtained by taking classical limit of \eqref{soft1} following the analysis given in \cite{G waves}.

To summarise in this section we have derived following conservation law :
\begin{align}
F^{[\log u/r^2]}_{rA}(\hat{x})|_{\mathcal{I}^+_-}=
F^{[\log r/r^2]}_{rA}(-\hat{x})|_{\mathcal{I}^-_+}.
\end{align}
This conservation law was shown to be equivalent to the 1-loop level $\log \omega$ soft theorem in \cite{1903.09133, 1912.10229}. 

\section{Subleading tail to the memory term}
In this section we will derive the subleading tail to the memory term. 

Let us first outline our steps. In the last section we studied the asymptotic field at $\mathcal{O} (e^3)$. The $\mathcal{O} (e^3)$ modes in the radiation arise due to acceleration of the charged particles under the long range electromagnetic force. This radiation backreacts on the particles. When we go to higher orders in $e$ we need to include the effect of this backreaction.
We show in Appendix \ref{lrf} that including the effect of the emitted radiation at $\mathcal{O} (e^3)$, the asymptotic forces $(F)$ that act on a scattered particle take following form \eqref{F22} :
$$ F(\tau) =e\ \sum_{m=2}^\infty\frac{c_m}{\tau^m}+e^3\ \sum_{n=3}^\infty d_n\frac{\log\tau}{\tau^n}.$$
The leading order $\frac{1}{\tau^2}$-term is the $\mathcal{O}(e)$ term that we had studied earlier. It gives rise to logarithmic correction to the trajectory we had discussed in \eqref{x1}. This term does not get corrected at $\mathcal{O} (e^3)$. At $\mathcal{O}(e^3)$, the leading term is the $\frac{\log\tau}{\tau^3}$-term. Following the analysis in Appendix \ref{lrf}, it is clear that the coefficient of this term is fixed by the coefficient of the $\frac{1}{\tau^2}$-term. It depends only on the charges and asymptotic velocities of the scattering particles. As we see in \eqref{FD}, the higher order moments contribute at $n>3$. Thus the $n>3$ modes depend on the charge distribution of the scattering objects.

As a result of $\frac{\log\tau}{\tau^3}$-force, the asymptotic trajectories of the particles are corrected to :
\be x^\mu_i=V^\mu_i \tau +c^\mu_i \log \tau + d_i+\eta_if_{i\sigma}\frac{\log\tau}{\tau}, \label{x22} \ee
where using \eqref{f}\footnote{Earlier, I had missed the second part of the first term i.e. the one proportional to $p_j.c_i\ p_i.p_j$. I am extremely thankful to Biswajit Sahoo for pointing this out in \cite{Bis}.}
\begin{align} f^\mu_i = -\sum_{\substack{j=m+1,\\ i\neq j}}^n m_im_j^2\frac{Q_iQ_j}{2} \Big[ 3(m^2_jp_i.c_j-p_j.c_i\ p_i.p_j)\  \frac{( p_i.p_j\ p_i^\mu+ m_i^2 p^\mu_{j})}{[(p_i.p_j)^2-m_i^2m_j^2]^{5/2}}+ \frac{[ p_i.p_j\ c^\mu_{i}-( p_i.p_j\ c^\mu_{j}-p_i.c_j\  p^\mu_{j})]}{[(p_i.p_j)^2-m_i^2m_j^2]^{3/2}} \Big].\label{f2}\end{align}
In \eqref{x22}, we have ignored $\mathcal{O}(\frac{1}{\tau})$ corrections that are not relevant for our analysis. The $f_{i\sigma}$ term is a result of backreaction of the radiation. Particles will in turn radiate because of this second order deviation in trajectories. This radiation is at $\mathcal{O}(e^5)$. Including these $\mathcal{O}(e^5)$ effects, we will obtain the subleading tail to the memory term. So let us find the resultant correction to radiative field due to the $\mathcal{O}(\frac{\log\tau}{\tau})$ correction to the trajectory. The field generated by $i^{th}$ outgoing particle is given using the retarded Green function : 
\begin{align}
A^{(i)}_\sigma(x)
&=\frac{1}{2\pi }\int d^4x'\ \delta(\ (x-x')^2)\ j^{(i)}_\sigma(x')\  \Theta(t-t').
\end{align}
Here we need to use the modified current calculated using the corrected trajectory given in \eqref{x22}.
$$j^{(i)}_\sigma(x') = \int d\tau \ e_i \big[V_{i\sigma}+\frac{c_{i\sigma}}{\tau}-f_{i\sigma}\frac{\log\tau}{\tau^2}\big]\  \delta^4(x'-x_i)\ \Theta(\tau-T).$$
We have to solve $(x-x')^2=0$ to second order in coupling $e$. The solution is given in \eqref{tau22} in Appendix A. Next we will use :
\begin{align}
\delta(\ (x-x')^2)
&= \frac{\delta(\tau-\tau_2)}{|2\tau+2V_i.(x-d_i)+\frac{2(x-d_i).c_i}{\tau}-2f_i.(x-d_i)\frac{\log\tau}{\tau^2}+2\frac{f_i.(x-d_i)}{\tau^2}-2c_i^2\frac{\log \tau}{\tau}| }.\label{A22}
\end{align}
To study the subleading correction to the memory term, we need the leading $\frac{1}{r}$ term in $A_\sigma$. So, we will ignore the terms in the solution that contribute at $\mathcal{O}(\frac{1}{r^2})$ or higher. The solution to the asymptotic field at this order turns out to be :
\begin{align}
A^{(i)}_\sigma(x)
&=\frac{1}{4\pi }  \frac{\Theta(\tau_2)\ e_i \big[V_{i\sigma}+\frac{c_{i\sigma}}{\tau_2}-f_{i\sigma}\frac{\log\tau_2}{\tau^2_2}\big] }{|V_i.x+\frac{x.c_i}{\tau_2}-f_i.x\frac{\log\tau_2}{\tau_2^2}+\frac{f_i.x}{\tau_2^2}+\mathcal{O}(r^0)|} \label{A2}.
\end{align}
Next it only remains to subtitute the value of $\tau_2$. We get it from \eqref{tau2} :
\begin{align}
\tau_2|_{\mathcal{I}^+}&=-\frac{u}{q.V_i}- \ \frac{q.c_i}{q.V_i} \ \log u+ q.f_i\ \frac{ \log u}{u}- \frac{(q.c_i)^2}{q.V_i}\ \frac{ \log u}{u}+\mathcal{O}(1).\nn
\end{align}
We will substitute above solution of $\tau_2$ in \eqref{A2} to obtain the leading $\frac{1}{r}$ term in $A_\sigma$.
\begin{align}
A^{(i)}_\sigma(x)|_{\mathcal{I}^+}
&= -\frac{e_i \Theta(\tau_2)}{4\pi r} \Big[ \frac{V_{i\sigma}}{ q.V_i}  \big[1+\frac{q.c_i}{u}-(q.c_i)^2\frac{\log u}{u^2}+{q.f_i}\ {q.V_i}\frac{\log u}{u^2} \big]    -\frac{c_{i\sigma}}{u} \big[1 -{q.c_i}\frac{\log u}{u}   \big]  -f_{i\sigma}\ q.V_i\frac{\log u}{u^2} \Big].
\end{align}
To get the full asymptotic field, we just need to sum over contributions from all particles. ($Q_i,p_i$ have been defined in \eqref{eta}).
\begin{align}
A_\sigma(x)|_{\mathcal{I}^+}
&=-\frac{1 }{4\pi r}\Big[\sum_{i=n'+1}^n  \Theta(u-T)\ \frac{Q_ip_{i\sigma}}{q.p_i}\ -\  \sum_{i=1}^{n'} \Theta(-u-T)\ \frac{Q_ip_{i\sigma}}{q.p_i}\ \Big]\nn\\
&+ \frac{1}{4\pi ru}\ \Big[\sum_{i=n'+1}^{n} \Theta(u-T) \ Q_i\big[c_{i\sigma}-p_{i\sigma}\frac{q.c_i}{q.p_i}\big]\ -\ \sum_{i=1}^{n'} \Theta(-u-T) \ Q_i\big[c_{i\sigma}-p_{i\sigma}\frac{q.c_i^-}{q.p_i}\big]\ \Big] \nn\\
&-\frac{1 }{4\pi r}\ \frac{\log u}{u^2}\ \Big[ \sum_{i=n'+1}^n \ \Theta(u-T)Q_i\big[q.c_i\ c_{i\sigma}-p_{i\sigma}\frac{(q.c_i)^2}{q.p_i} +\frac{p_{i\sigma}}{m_i}{f_i.q}-\frac{q.p_i}{m_i}f_{i\sigma}\big]\nn\\
&-\sum_{i=1}^{n'} \ \Theta(-u-T)\  Q_i\big[ q.c_i\ c_{i\sigma}-p_{i\sigma}\frac{(q.c_i)^2}{q.p_i}+\frac{p_{i\sigma}}{m_i}{f_i.q}-\frac{q.p_i}{m_i}f_{i\sigma}\big]\ \Big]+...\ . \label{Alogu}
\end{align}
%
%
This is the expression for late time radiative field at large $u$. We have already discussed the leading term which is $\mathcal{O}(u^0)$. It gives rise to the so called memory effect \cite{mem4, Garfinkle}. The $\mathcal{O}(\frac{1}{u})$ term is the leading order tail to the memory term \cite{log mem,log mem em}. From above expression, we see that the subleading tail to the $\mathcal{O}(u^0)$ memory term is $\mathcal{O}(\frac{\log u}{u^2})$. These tail terms can be observed directly by observing velocity profile of a test particle placed at large $r$ and large $u$. The transverse velocity of a test particle $V^A$ at late times settles down to the final value determined by $u^0$ memory term as we had discussed in section 2.2. The approach to this final value is controlled by these tail terms.

The subleading tail derived in \eqref{Alogu} is the result of this section. Like the memory and the tail terms, the subleading tail depends only on asymptotic momenta and charges of the interacting particles. It is insensitive to the details of the scattering. In Appendix \ref{int}, we also see that it is insensitive to the non-minimal couplings as these interactions fall off faster. This is the same reason why spin of the particle does not play any role and why higher derivative corrections to Lagrangian do not affect this term. Therefore these leading logs are universal for any U(1) gauge invariant theory. \footnote{The subleading tail term will change if we modify the coefficients of $\frac{1}{\tau^2}$ and $\frac{\log\tau}{\tau^3}$ terms. We can think of only one physical scenario where this happens : if add gravity.} This term appears at $\mathcal{O}(e^5)$. We have checked that it is not corrected at $\mathcal{O}(e^7)$. We expect that the coefficient of this term stays uncorrected to all orders in $e$. Because of its universal nature it seems that the subleading tail is also tied to a soft theorem. 

Let us briefly discuss the frequency space radiative field. We rewrite \eqref{Alogu} as :
\begin{align}
A_\mu|_{\mathcal{I}^+}=\frac{1}{4\pi r} \Big[\ a_\mu^\pm u^0\ +\ \frac{[b^{(0)}_\mu]^\pm}{u}\ +\ [b^{(1)}_\mu]^\pm \frac{\log u}{u^2}\ \Big] +...,\ \ \  u\rightarrow\pm\infty.\label{falloff2}
\end{align}
The large-$u$ fall offs given in \eqref{falloff2} fix the small frequency behaviour of the Fourier transformed function. The frequency space radiative field has following behaviour at small $\omega$ \cite{log mem em} :
\begin{align}
 \tilde{A}_\mu(\omega) =\frac{e^{i\omega r}}{4\pi ir}\Big[\ \frac{1}{\omega}\ [a_\mu^+-a_\mu^-]\ -\ i\big[\ [b^{(0)}_\mu]_\mu^+-[b^{(0)}_\mu]^-\big]\ \ \log\omega -\ \frac{1}{2}\big[\ [b^{(1)}_\mu]_\mu^+-[b^{(1)}_\mu]^-\big]\ \omega(\log\omega)^2+...\Big] \ \ \text{as }\omega\rightarrow 0.
\end{align}
From \eqref{Alogu} we can read off the coefficient of the $\omega(\log\omega)^2$-term :\footnote{It is tempting to compare $S_2$ with the tree level subsubleading soft factor. From \cite{Hamada Shiu, soft inf} :
\begin{align}
\mathcal{S}_{\text{tree}}^2 = \frac{1}{2}\sum_{i=1}^n \Big[\ Q_i  \ [\frac{\epsilon.p_{i}}{ q.p_i} (q.\p_i)-\epsilon.\p_{i} ]\ (q.\p_i) \Big]\mathcal{A}_{n}+\epsilon^\mu q^\nu\ {A}_{\mu\nu}(p_i)\ . \label{tree}
\end{align}
See discussion around \eqref{softt} for details.}
\begin{align}
S_2&=-\frac{1}{2}\sum_{i=1}^nQ_i \Big[ q.c_i\ \big[p_{i}.\epsilon\frac{q.c_{i}}{p_i.q}-c_{i}.{\epsilon}\big]-\epsilon_\mu q_\nu \ [p_i^\mu f_i^\nu-p_i^\nu f_i^\mu]\ \Big].\label{subclass}
\end{align}
This analysis hints at following universal term in the soft expansion for 2-loop amplitudes :
\begin{align}
\mathcal{A}_{n+1}(p_i,k)&=\frac{{S}_0}{\omega}\ \mathcal{A}_n(p_i)\  + \ S_{\text{1-loop}}\log{\omega}\ \mathcal{A}_n(p_i)\ + \ S_{\text{2-loop}}\ \omega(\log\omega)^2\ \mathcal{A}_n(p_i)\ + ...\ .\label{soft2}
\end{align}
Here, $S_2$ would be the classical limit of $S_{\text{2-loop}}$. An interesting point to note is that $S_2$ is made up of two gauge invariant terms, the first term in the bigger square bracket is gauge invariant by itself and so is the second term. This point becomes important later on.

\section{$m=2$ conservation law}
We have calculated the subleading tail to the memory term which turns out to be universal. Therefore we suspect that it is related to classical limit of a new soft theorem. This possibility needs to be explored further. As a first step in this direction, we show existence of an asymptotic conservation law that is associated to the subleading tail term. Let us briefly describe our results. When we include the $\mathcal{O}(e^5)$ terms, the asymptotic expansion for $F_{rA}$ takes following form around future null infinity : 
\be
F^{[1/r^3]}_{rA}|_{u \rightarrow -\infty} =  u^2\ F^{[u^2/r^3]}_{rA}(\hat{x})\ +u\log u\ F^{[u\log u/r^3]}_{rA}(\hat{x})+ (\log u)^2\ F^{[(\log u)^2/r^3]}_{rA}(\hat{x})\ + ... \  . \ee
Expansion around the past null infinity is given by : 
\be
F_{rA}|_{v \rightarrow \infty} = \frac{\log r}{r^2}\ [v^0 \ F^{\log r/r^2]}_{rA}(\hat{x})\ +...]+ \frac{(\log r)^2}{r^3}\ [\ v^0\ F^{[{(\log r)^2}/{r^3}]}_{rA}(\hat{x})\  +...] + \mathcal{O}(\frac{1}{r^2})\ .\ee
 In this section, we show that a generic classical scattering process obeys following conservation law : 
\begin{align}
F^{[(\log u)^2/r^3]}_{rA}(\hat{x})|_{\mathcal{I}^+_-}=-
F^{[(\log r)^2/r^3]}_{rA}(-\hat{x})|_{\mathcal{I}^-_+}.\label{2cons}
\end{align}

Next we will calculate the asymptotic field configuration to prove above conservation law. A reader who is not interested in the details of the derivation can jump ahead to \eqref{2loop}. Let us go back to \eqref{A2} that included the relevant $\mathcal{O}(e^5)$ corrections. We need to retain the subleading corrections in $\frac{1}{r}$ since we need to obtain the $\mathcal{O}(\frac{\log u}{r^3})$ mode.
\begin{align}
A^{(i)}_\sigma(x)
&=\frac{1}{4\pi } \frac{ e_i \big[V_{i\sigma}+\frac{c_{i\sigma}}{\tau_2}-f_{i\sigma}\frac{\log\tau_2}{\tau_2^2}\big]\   \Theta(\tau_2)}{|\tau_2+V_i.(x-d_i)+\frac{(x-d_i).c_i}{\tau_2}-f_i.x\frac{\log\tau_2}{\tau_2^2}+\frac{f_i.x}{\tau_2^2}-c_i^2\frac{\log \tau_2}{\tau_2}| }.\label{A23}
\end{align}
We recall that $\tau_2$ is given in \eqref{tau22}. At ${\mathcal{I}^+}$, the charge is expected to be defined in terms of $\frac{(\log u)^2}{r^3}$-mode of $A_\sigma$. Using \eqref{tau2}, we get following expansion around ${\mathcal{I}^+}$ :
$$\frac{1}{\tau_2}=-\frac{q.V_i}{u}\Big[1+\mathcal{O}(\frac{\log u}{u})\Big].$$
Let us dicuss the second term in the numerator i.e. $\frac{ e_i}{|\tau_2+V_i.x+...|}\frac{c_{i\sigma}}{\tau_2}$. The expansion of this term is of following form : $$\frac{1}{ur}\Big[1+\frac{\log u}{u}+...+\frac{1}{r}[u+\log u +u^0]+\frac{1}{r^2}[u^2+u\log u +(\log u)^2+...]+...\Big].$$ 
Thus the second term in the numerator does not have $\frac{(\log u)^2}{r^3}$ mode. Next we turn to the last term in the numerator that can be expanded as follows 
 $$\frac{\log u}{u^2r}\Big[1+\frac{1}{r}[u+ u^0]+\frac{1}{r^2}[u^2+...]+...\Big].$$ 
Thus the last term in the numerator also does not contribute to $\frac{(\log u)^2}{r^3}$. Similarly, last four terms in the denominator do not contribute to $\frac{(\log u)^2}{r^3}$ and hence are irrelevant for subsequent analysis. So, we are left with following terms in \eqref{A23} :
\begin{align}
A^{(i)}_\sigma(x)
&\sim\frac{1}{4\pi } \frac{ e_i V_{i\sigma}  \Theta(\tau_2)}{|\tau_2+V_i.(x-d_i)| }.
\end{align}
We have used '$\sim$' instead of '$=$' as we are ignoring certain terms in $A_\sigma$ that do not contribute to the charge given in \eqref{2cons}. Next we need to substitute the value of $\tau_2$. The full expression is given in \eqref{tau22}. Let us retain only the logarithmic terms in $\tau_2$ that are of relevance to us. We get :
\be 
A^{(i)}_\sigma(x)|_{\mathcal{I}^+}\sim\frac{1}{4\pi } \frac{ e_i \ V_{i\sigma}\ \Theta(u)}{X[1-2x.c_i \frac{\log u}{X^2} + c_i^2 \frac{(\log u)^2}{X^2}]^{1/2}}.\label{A24}
\ee
%
We recall that $X=[\ (V_i.x-V_i.d_i)^2 +(x-d_i)^2\ ]^{1/2}$. Hence, the limiting value of $X$ is $X|_{\mathcal{I}^+}=-rq.V_i+\mathcal{O}(r^0)$.  Substituting the value of $X$ the coefficient of $\frac{(\log u)^2}{r^3}$ comes out to be : 
\be 
A^{(i)}_\sigma(x)|_{\mathcal{I}^+}\sim-\frac{1}{8\pi }\frac{(\log u)^2}{r^3} \frac{V_{i\sigma}}{q.V_i}\   \Big[3\frac{(q.c_i)^2}{ (q.V_i)^4}\ -\frac{c_i^2}{(q.V_i )^2}\Big].
\ee
Next we need to sum over the contributions from all particles. Let us write down the coefficient at $\mathcal{I}^+_-$. The contribution around $\mathcal{I}^+_-$ is from the incoming particles, thus we get :
\begin{align}
A^{[(\log u)^2/{r^3}]}_\sigma(\hat{x})|_{\mathcal{I}^+_-}&=-\sum_{i=1}^{n'}\frac{e_i}{8\pi }   \frac{V_{i\sigma}}{q.V_i}\   \Big[3\frac{(q.c_i)^2}{ (q.V_i)^4}\ -\frac{c_i^2}{(q.V_i )^2}\Big].
\end{align}
We just need to transform co-ordinates to go to $F_{rA}$. Thus :
\begin{align}
F_{rA}^{[(\log u)^2/{r^3}]}(\hat{x})|_{u\rightarrow -\infty}&=  \sum_{i=1}^{n'}\frac{e_i}{4\pi }\    \frac{3(q.c_i)}{ (q.V_i)^5}\ q^\mu (\p_A q^\nu )   \ [V_{i\nu} c_{i\mu}-V_{i\mu} c_{i\nu}].\label{Ff}
\end{align}
Let us carry out the corresponding calculation at past null infinity. At ${\mathcal{I}^-}$, the term of our interest is the $\frac{(\log r)^2}{r^3}$-mode of $A_\sigma$. Following earlier logic and using \eqref{t2p} analogous to \eqref{A24},  we get following expression at past null infinity we get :
$$A^{(i)}_\sigma(x)|_{\mathcal{I}^-}\sim \frac{ e_i \ V_{i\sigma}\ }{X[1-2x.c_i \frac{\log r}{X^2} + c_i^2 \frac{(\log r)^2}{X^2}]^{1/2}}.$$ 
Using \eqref{qbar} in $X=[\ (V_i.x-V_i.d_i)^2 +(x-d_i)^2\ ]^{1/2}$, the limiting value at past null infinity turns out to be $X|_{\mathcal{I}^-}=r\bar{q}.V_i+\mathcal{O}(r^0)$. Expanding above expression, we obtain the $\frac{(\log r)^2}{r^3}$ term in $A_\sigma$ :
\begin{align}
A^{[(\log r)^2/{r^3}]}_\sigma(x)|_{\mathcal{I}^-}
=   \sum_{i=1}^{n'} \frac{e_i}{8\pi } \frac{V_{i\sigma}}{V_i.\bar{q}}\   \Big[\ 3 \frac{(\bar{q}.c_i)^2}{ (V_i.\bar{q})^4}\ -\frac{c_i^2}{(V_i.\bar{q})^2}\ \Big]
. \ 
\end{align}
Next we just need to use appropiate co-ordinate transformations to arrive at $F_{rA}$. We get :
\begin{align}
F_{rA}^{[(\log r)^2/r^3]}(\hat{x})|_{v\rightarrow \infty}
&=\sum_{i=1}^{n'}\frac{e_i}{4\pi }\   \frac{3(\bar{q}.c_i)}{ (\bar{q}.V_i)^5}\  \bar{q}^\mu  \ (\p_A\bar{q}^{\nu})\ [V_{i\mu} c_{i\nu}-V_{i\nu} c_{i\mu}]. \label{Fp}
\end{align}
Using \eqref{Ff} and \eqref{Fp}, we can write down the conservation law for these modes :
\begin{align}
F^{[(\log u)^2/r^3]}_{rA}(\hat{x})|_{\mathcal{I}^+_-}=-
F^{[(\log r)^2/r^3}_{rA}(-\hat{x})|_{\mathcal{I}^-_+}.\label{2loop}
\end{align}
It is important to note the minus sign. The charges can be defined as $Q_{2}^+=\int d^2z \ Y^A\ F^{[(\log u)^2/r^3]}_{rA}(\hat{x})|_{\mathcal{I}^+_-}$ and $Q^-_{2}=\int d^2z \ Y^A\ F^{[(\log r)^2/r^3]}_{rA}(-\hat{x})|_{\mathcal{I}^-_+}$. 
We show in \eqref{201} in Appendix C that $\frac{(\log u)^2}{r^3}$ mode is related to the subleading tail in the late time radiative field. To summarise we have derived a conservation law \eqref{2loop} that is related to the subleading tail term we obtained in \eqref{Alogu}.

Let us briefly discuss what are the implications of this charge on the quantum amplitudes. We expect that the corresponding Ward identity for the $S$-matrix i.e. $[Q_2,S]=0$ would be related to the new soft theorem that we had discussed in \eqref{soft2}. Since the charge $Q_2$ involves $\mathcal{O}(e^5)$ modes, this $\omega(\log\omega)^2$ term in the soft expansion will appear at 2-loop order.

Next we need to highlight an important point about this charge. We recall the subsubleading classical soft factor $S_2$ we discussed in eq.\eqref{subclass} :
\begin{align}
S_2&=-\frac{1}{2}\sum_{i=1}^nQ_i \Big[ \frac{(q.c_i)}{ (q.p_i)}\   \big[\ p_{i}.\epsilon\ {q.c_{i}}-c_{i}.{\epsilon}\ {p_i.q}\ \big]-\frac{1}{m_i}[ \ p_{i}.\epsilon\ {q.f_{i}}\  -\ p_i.q\ \epsilon.f_i\ ]\ \Big].\label{1}
\end{align}
Let us go back to the expression of the charge $Q_2$. The charges have both soft modes as well as hard modes. The future charge includes the contribution of the soft modes which is given by $Q^s_2 = F_{rz}^{[(\log u)^2/{r^3}]} |_{\mathcal{I}^+_+}- F_{rz}^{[(\log u)^2/{r^3}]}|_{\mathcal{I}^+_-}$. Here we have chosen $Y^{\bar{z}}=0$. Using \eqref{Ff}, we obtain the explicit expression :
\begin{align}
Q^s_2&=  \sum_{i=1}^n\frac{Q_i}{4\pi }\    \frac{3(q.c_i)}{ (q.p_i)^5}\ q^\mu (\p_z q^\nu )   \ [p_{i\nu} c_{i\mu}-p_{i\mu} c_{i\nu}],\nn\\
&=  \sum_{i=1}^n\frac{Q_i}{4\pi }\    \frac{3(q.c_i)}{ (q.p_i)^5}\   \big[\ p_{i}.\epsilon^+\ {q.c_{i}}-c_{i}.{\epsilon^+}\ {p_i.q}\ \big].\label{0}
\end{align}
In the last line, we have used the basis given in \eqref{pol} for polariation vectors. Comparing eq. \eqref{1} to eq. \eqref{0} we see that the second half of the soft factor i.e. $\frac{1}{2}\large{\sum_{i=1}^n\frac{Q_i}{m_i}[ \ p_{i}.\epsilon\ {q.f_{i}}\ -\ p_i.q\ \epsilon.f_i]}$ does not contribute to the $m=2$ charge. This implies that this part of the soft factor is not controlled by the asymptotic charge $Q_2$. We had already discussed that this term is gauge invariant by itself. We will call $\frac{1}{2}\large{\sum_{i=1}^n\frac{Q_i}{m_i}[ \ p_{i}.\epsilon\ {q.f_{i}}\ -\ p_i.q\ \epsilon.f_i]}$ as a remainder term since it is not controlled by the asymptotic charge\footnote{ It is again tempting to compare with the tree level case where the conservation law given in \cite{infinite asym} produces only first gauge invariant term in the soft factor given in \eqref{tree}. The second gauge invariant term whose structure is similar to our remainder term is not controlled by any charge. }. But intriguingly the remainder term is also universal. We believe that the remainder term need to be understood better.

\section{Proposal for conservation laws for general $m$}
Based on the $m=1,2$ conservation laws, we propose that there exists a conservation law for every $m$ given by :
\begin{align}
F^{[(\log u)^m/r^{m+1}]}_{rA}(\hat{x})|_{\mathcal{I}^+_-}=(-1)^{m+1}\ 
F^{[(\log r)^m/r^{m+1}]}_{rA}(-\hat{x})|_{\mathcal{I}^-_+}.\label{ncons}
\end{align}
Thus, we expect that classical electromagnetism admits a hierarchy of infinite number of conservation laws. The $m^{th}$ level future charge is defined as $Q^+_m=\int d^2z \ Y^A_m(\hat{x})\ F^{[(\log u)^m/r^{m+1}]}_{rA}(\hat{x})|_{\mathcal{I}^+_-}$ and the past charge is $Q^-_m=\int d^2z \ Y^A_m(-\hat{x})\ F^{[(\log r)^m/r^{m+1}]}_{rA}(-\hat{x})|_{\mathcal{I}^-_+}$. These charges are $\mathcal{O}(e^{2m+1})$.

Let us discuss the corresponding universal terms in the late time radiation. When we include the effect of long range forces on the trajectory, the full correction to the trajectory is of the form :
\be x'^\sigma_i = V_{i\sigma}\tau+c_{i\sigma}\log\tau+d_{i\sigma}+\sum_{\substack{m\leq n\\ m=0,n=1}}^\infty{c^{(m,n)}_{i\sigma}}\frac{(\log{\tau})^m}{\tau^n},\label{a1}\ee
where $c^{(m,n)}_{i\sigma}$'s typically admit a series expansion in the coupling $e$. The leading logarithmic terms are fixed in terms of the coefficient of $\frac{1}{\tau^2}$ term in electromagnetic field, hence they depend only on the charges and the asymptotic momenta of the scattering particles. Subleading terms depend on details of scattering like impact parameters ($d_i$) and also get contribution from higher order moments of the scattering bodies. So, not all the $c^{(m,n)}_{i\sigma}$'s are universal. Substituting the corrected trajectories of \eqref{a1} in the solution :
\begin{align}
A_\sigma(x)
&=\frac{1}{2\pi }\int d^4x'\ \delta([x-x']^2)\ j_\sigma(x')\  \Theta(t-t') .
\end{align}
we get following profile for the late time radiative field :
\begin{align}
A_\mu=\frac{1}{4\pi r} \Big[\ a_\mu^\pm u^0\ +\ \frac{b_\mu^\pm}{u}\ +\ \sum_{\substack{m< n\\ m=0,n=2}}^\infty [g^{m,n}_\mu]^\pm \frac{(\log u)^m}{u^n}\ \Big] ,\ \ \  u\rightarrow\pm\infty.\label{a4}
\end{align}
As we discussed earlier typically $[g^{m,n}_\mu]^\pm$'s are not universal. Now we use the fact that $\frac{(\log u)^n}{r^{n+1}}$-term in $F_{rA}$ in fixed by $\frac{(\log u)^{m-1}}{ru^m}$- term in $A_\sigma$ by Maxwells equations. Hence we expect that $\frac{(\log u)^{m-1}}{ru^m}$ tail terms in $A_\sigma$ must be universal as they are related to asymptotic charge. To summarise we consider following class of terms in the late time radiation :
\begin{align}
A_\mu=\frac{1}{4\pi r} \Big[\ a_\mu^\pm u^0\  +\ \sum_{\substack{m=1}}^\infty [b^{(m-1)}_\mu]^\pm \frac{(\log u)^{m-1}}{u^m}\ +...\Big] ,\ \ \  u\rightarrow\pm\infty, \label{Atail}
\end{align}
where $[b^{(m-1)}_\mu]^\pm$ is equal to $[g^{m-1,m}_\mu]^\pm$ of \eqref{a4} for $m>1$ and $[b^{(m-1)}_\mu]^\pm$ is equal to $b_\mu^\pm$ of \eqref{a4} for $m=1$. We expect that above terms are controlled by soft theorems. Here $b^{(0)}_{\mu}$ is the first order tail to the memory \cite{log mem}. $b^{(1)}_\mu$ is the subleading tail we calculated in section 4.

It should be noted that the universal terms in the late time field arise from following class of corrections to the equation of trajectory  given in \eqref{a1} :
\be x'^\sigma_i = V_{i\sigma}\tau+c_{i\sigma}\log\tau+d_{i\sigma}+\sum_{m=1}^\infty{(\eta_i)^mf^{(m)}_{i\sigma}}\frac{(\log{\tau})^m}{\tau^m}+...\ ,\label{a2}\ee
here we have defined $(\eta_i)^mf^{(m)}_{i\sigma}=c_{i\sigma}^{m,m}$. $\eta_i=1 (-1)$ for outgoing (incoming) particles. So that $f^{(1)}_{i\sigma}$is the $f_{i\sigma}$ that appeared in the subleading tail calculated in section 4. 

Using the technique developed in \cite{log mem em}, it can be shown that the fourier transform of \eqref{Atail} has following behaviour near $\omega\rightarrow 0$ :
\begin{align}
 \tilde{A}^\mu(\omega) =\frac{e^{i\omega r}}{4\pi ir}\Big[\ \frac{S_0^\mu}{\omega}\ [a_\mu^+-a_\mu^-]\ +\sum_{m=1}^\infty S_m^\mu\ \omega^{m-1}(\log\omega)^m +...\Big].\label{Atil}
\end{align}
The soft factors are given by $S_m=\epsilon_\mu S^\mu _m$. Based on the $m=1,2$ cases, we predict that the form of the classical soft factor for any $m$ is given by :
\begin{align}
S_m= \frac{ (i)^{m}}{m!}\sum_{i=1}^n \Big[\ Q_i  \ [\frac{\epsilon.p_{i}}{ q.p_i} (q.c_i)-\epsilon.c_{i} ]\ (q.c_i)^{m-1} +\epsilon^\mu q^{\nu_1}\cdots q^{\nu_{m-1}}\ \mathcal{F}_{i\mu\nu_1\cdots \nu_{m-1}}\ \Big], \label{softf}
\end{align}
where
\be\ \mathcal{F}_{i\mu\nu_1\cdots \nu_{m-1}}=(m-1)\sum_{m'=2}^m Q_i\ c_{i\nu_1}\cdots c_{i\nu_{m-m'}}\frac{p_{i\nu_{m-m'+1}}}{m_i}\cdots \frac{p_{i[\nu_{m-1}}}{m_i}\ f^{(m'-1)}_{i\mu]}.\label{FF}\ee
$f^{(m)}_{i\mu}$'s have been defined in \eqref{a2}. The indices have been enclosed within square brackets to denote that $\mu$ has to be antisymmetrised with $\nu_{m-1}$\footnote{ 
\be p_{i[\nu_{m-1}}\ f^{(m'-1)}_{i\mu]}=\ [f^{(m'-1)}_{i\mu}\ p_{i\nu_{m-1}}-p_{i\mu} f^{(m'-1)}_{i\nu_{m-1}}\ ].\nn\ee}. 

Let us make some comments about our proposal. We first recall the $m=2$ case i.e. eq.\eqref{subclass}.
\begin{align}
S_2=-\frac{1}{2} \sum_{i=1}^n \Big[\ Q_i  \ [\frac{\epsilon.p_{i}}{ q.p_i} (q.c_i)-\epsilon.c_{i} ]\ (q.c_i) +\ \epsilon_\mu q_{\nu}\ \mathcal{F}_i^{\mu\nu}\ \Big].\label{s2}
\end{align}
At the end of section 5,, we showed that the $m=2$ asymptotic conservation law given in \eqref{2cons} controls the first term in \eqref{s2}. The $\mathcal{F}_{i\mu\nu}$-term given by $\frac{Q_i}{m_i}[f^{(1)}_{i\mu}\ p_{i\nu}-p_{i\mu} f^{(1)}_{i\nu}\ ]$ is not by fixed by the asymptotic charge given in \eqref{2cons}. Hence we classified this term as a remainder term. We expect this to be true for all $m$'s. Hence, we will call the $\mathcal{F}$-term in \eqref{softf} as a remainder term since it is not controlled by the $Q_m$-conservation law given in \eqref{ncons}. We have not given the explicit expression of the remainder term, it can be obtained by calcualating the $f^{(m)}_\mu$ corrections to the trajectory. This term is also expected to be universal for all $m$'s. The first term in \eqref{softf} is fixed by the asymptotic charges in \eqref{ncons} and gauge invariance.

It is interesting to compare \eqref{softf} with the tree level coefficients that appear in soft expansion of amplitudes. For tree level amplitudes the soft expansion is given by $\mathcal{A}_{n+1}\sim \frac{S_0}{\omega}+\sum_{m=1}^\infty \omega^m\ \mathcal{S}_{\text{tree}}^m .$ From \cite{Hamada Shiu, soft inf} :
\begin{align}
\mathcal{S}_{\text{tree}}^m = \frac{1}{m!}\sum_{i=1}^n \Big[\ Q_i  \ [\frac{\epsilon.p_{i}}{ q.p_i} (q.\p_i)-\epsilon.\p_{i} ]\ (q.\p_i)^{m-1} \Big]\mathcal{A}_{n}+\epsilon^\mu q^{\nu_1}\cdots q^{\nu_{m-1}}\ {A}_{\mu\nu_1\cdots \nu_{m-1}}\ . \label{softt}
\end{align}
The first term is universal and is fixed by asymptotic charge \cite{infinite asym}. $A^{\mu\nu_1\cdots \nu_{m-1}}$ is an arbitrary tensor antisymmetric in $\mu$ and every $\nu_i$. This term is called remainder term. It depends on hard momenta and its form cannot be fixed by symmetry. Factorisation into lower point amplitude is not guaranteed for this term. The structure of the first term of \eqref{softt} is very similar to that of the first term of \eqref{softf}. We note that we can obtain the the first term in \eqref{softf} replacing $\p_k^{\nu}\rightarrow i c_k^\nu $ in the first term of \eqref{softt}. The structure of the remainder terms in the two expressions is also similar. 

\section{Summary}

Soft theorems have been studied extensively. Interestingly, the leading soft photon theorem controls the velocity kick experienced by a test charge making the leading soft factor directly observable in experiments. 

In this paper we studied the subleading tail to the leading order memory term in the late time radiation that appears at $\mathcal{O}(e^5)$. We obtained the expression of this term for a general scattering process in \eqref{Alogu}. This term depends on the asymptotic momenta and charges of the scattering particles. It is insensititve to other aspects of the interacting particles like spin or charge distribution as discussed in Appendix \ref{int}. It is also independent of the details of the scattering. It is fixed completely by the minimal coupling of particles with the gauge field hence insensitive to U(1) invariant non-minimal couplings or higher derivative corrections. 

We showed that there exists as an asymptotic conservation law ($m=2$ given in \eqref{2loop})  that is related to the subleading tail term. This charge fixes only a part of the coefficient of the subleading tail. The rest of the part is gauge invariant by itself and we call it 'remainder' term. We also proved the asymptotic conservation law ($m=1$ given in \eqref{1loop}) that controls the leading tail to the memory term. 

These are the three main results of this paper. We expect that the subleading tail term and the $Q_1, Q_2$ charges are not corrected by any higher order terms in $e$. Since our calculations are perturbative in $e$, so it hard to give a direct proof of this statement. It would be great to have a rigorous proof in the future.

Building on the $m=1,2$ cases, we have proposed that classical scattering processes satisfy an $\mathcal{O}(e^{2m+1})$ asymptotic conservation law for every $m$. We expect that this proposal for $Q_m$-conservation law can be proved for generic $m$ by incorporating the effects of long range force in the analysis of \cite{infinite asym}. In the classical theory, the asymptotic charges imply existence of universal terms in low energy radiation emitted in the scattering process. These terms would be seen as $m^{th}$ order tails in the late time radiative field. We expect that these tails also have universal structure that is independent of the details of the scattering. The coefficient of these tails is predicted to be given by \eqref{softf}. For $m\geq 2$ there are some remainder terms in these late time tails that are not captured by the asymptotic charges. An interesting question that arises here is to understand the universal nature of the remainder terms.

A natural question to ask is what is the implication of these charges for the quantum theory. We expect that these $Q_m$ charges imply existence of $m$-loop soft theorems for every $m$. But these soft theorems have not been studied in the literature before. So one needs to extend the calculations of \cite{Sen Sahoo} to higher loops and derive these $m$-loop soft theorems. We expect that loop amplitudes would admit soft expansion similar to \eqref{Atil}. A related question is to check that the classical limit of $m$-loop soft factors matches with our prediction given in \eqref{softf} for generic $m$.  \\

Several questions are in order about the conserved charges $Q_m$. 
\begin{itemize}
\item Most importantly, it needs to be checked if indeed all the $\lbrace Q_{m}, m\geq 1\rbrace$ charges are independent or if they are related. 
\item What is the underlying symmetry? Do these charges correspond to new kind of large gauge transformations?
\item For $m\geq 2$ there exist  remainder terms $\mathcal{F}_i^{\mu\nu_1\cdots \nu_{m-1}}$ that are not fixed by these $Q_m$ charges. These term too are expected to be universal for all $m$'s and depend only on the electric charges and the asymptotic momenta of scattering particles. Can we extend $Q_m$ such that they reproduce the entire soft factor including the remainder terms? 
\end{itemize}

\section{Acknowledgements}
I am deeply grateful to Alok Laddha for many insightful discussions. I am deeply thankful to Biswajit Sahoo for pointing out a term that I had missed in \eqref{f2} earlier. Finally I thank the people of India for their support to basic sciences.

\appendix

\section{Perturbative solution}\label{pert}

The Green function for d'Alembertian operator is $\delta([x-x']^2)$. We will find the solution of this delta function perturbatively in coupling $e$. Here, $x'^\mu(\tau)$ is the equation of trajectory that gets corrected as we go to higher orders in $e$. We will write down the perturbative solution for $\tau$  and see that it involves even powers of $e$.

At zeroth order, we have free particles :
$$ x'^\mu_i= V_i^\mu \tau + d_i.$$
Hence, the root of delta function $\delta([x-x']^2)$ is given by :
\begin{align}
\tau_0&=-V_i.(x-d_i)-\big[\ (V_i.x-V_i.d_i)^2 +(x-d_i)^2\ \big]^{1/2}.\label{tau00}
\end{align}
The sign of the square root has been chosen to ensure retarded boundary condition i.e. $\Theta(t-t'(\tau))$.
Now, let us study above expression in the limit $r\rightarrow \infty$ with $u$ finite. Thus, around $\mathcal{I}^+$, using \eqref{q} we get  :
\be \tau_0|_{\mathcal{I}^+}=-\frac{u}{q.V_i}+\mathcal{O}(1).\label{tau0f}\ee
Now we take $r\rightarrow \infty$ limit of \eqref{tau00} keeping $v$ finite, using \eqref{qbar}, we get :
\begin{align}
\tau_0|_{\mathcal{I}^-}&=-2r\ V_i.\bar{q}   +\mathcal{O}(1).\nn
\end{align}

Next we include the leading order effect of long range electromagnetic force. We know that the first order correction to the trajectory is given by \eqref{x1} :
$$ x'^\mu_i=V^\mu_i \ \tau +c^\mu_i \log \tau + d_i.$$
Using the corrected trajectory, the solution of delta function $\delta(|x-x'|^2)$ is given by :
\be\tau^2+2\tau V_i.(x-d_i) -(x-d_i)^2=-2(x-d_i).c_i\log\tau+c_i^2(\log \tau)^2. \label{10}\ee
We have used $V_i.c_i=0$. Noting that $c^\mu$ is $\mathcal{O}(e^2)$, the RHS of above equation can be treated as a perturbation. Hence we substitute the zeroth order solution \eqref{tau00} in RHS of \eqref{10} that leads to following equation for $\tau$ :
\be\tau^2+2\tau V_i.(x-d_i) -(x-d_i)^2=-2(x-d_i).c_i\log\tau_0. \ee
We ignored the $c_i^2$ term as it is $\mathcal{O}(e^4)$. Now, above equation is just a quadratic equation in $\tau$ and the solution is given by :
\begin{align}
\tau_1&=-V_i.(x-d_i)-\big[\ (V_i.x-V_i.d_i)^2 +(x-d_i)^2\ -2(x-d_i).c_i\log\tau_0\ \big]^{1/2}.\label{tau1}
\end{align}
We have used a subscript 1 to denote that it includes the first order perturbative effects. We can expand the squareroot to $\mathcal{O}(e^2)$ : 
\begin{align}
\tau_1&=-V_i.(x-d_i)-\big[\ (V_i.x-V_i.d_i)^2 +(x-d_i)^2\big]^{1/2}\ +\frac{(x-d_i).c_i}{X}\log\tau_0\ .\label{tau11}
\end{align}
Here, we have defined $X=[\ (V_i.x-V_i.d_i)^2 +(x-d_i)^2\ ]^{1/2}$ and $\tau_0$ is given in \eqref{tau00}. Thus, the first order solution is the zeroth order solution plus a perturbation :
\begin{align}
\tau_1&=\tau_0 +\frac{(x-d_i).c_i}{X}\log\tau_0\ .\label{tau11}
\end{align}
 Expanding around $\mathcal{I}^+$, we get :
\begin{align}
\tau_1|_{\mathcal{I}^+}&=-\frac{u}{q.V_i}- \ \frac{q.c_i}{q.V_i} \ \log u+\mathcal{O}(1).\label{t1f}
\end{align}
Thus, in $u\rightarrow \pm\infty$ limit, the correction to $\tau_0$ is suppressed by $\frac{\log u}{u}$ in addition to the suppression due to $e^2$ factor.
Expanding \eqref{tau11} around $\mathcal{I}^-$, we get :
\begin{align}
\tau_1|_{\mathcal{I}^-}&=-2r\ V_i.\bar{q}+ \  \frac{\bar{q}.c_i}{V_i.\bar{q}} \ \log r+\mathcal{O}(1). \label{t1p}
\end{align}

\subsection{Second order in perturbation}
Let us repeat above steps with second order efects of long range forces. The trajectory is corrected to \eqref{x2} where $f_i\sim \mathcal{O}(e^4)$ :
$$x'_\sigma = V_{i\sigma}\tau+{c_{i\sigma}}\log{\tau}+d_{i\sigma}+f_{i\sigma}\frac{\log\tau}{\tau}.$$
Hence at $\mathcal{O}(e^4)$, $\delta(|x-x'|^2)$ implies following equation for $\tau$ :
\be\tau^2+2\tau V_i.(x-d_i) -(x-d_i)^2=-2(x-d_i).c_i\log\tau_1-2(x-d_i).f_i\frac{\log\tau_1}{\tau_1}+c_i^2(\log \tau_1)^2. \label{2}\ee
 We have used the fact that $V_i.c_i=V_i.f_i=0$.
Here, we have substituted the corrected solution \eqref{tau1} for the terms in the RHS. The second order solution is  :
\begin{align}
\tau_2&=-V_i.(x-d_i)-\big[\ (V_i.x-V_i.d_i)^2 +(x-d_i)^2\ -2(x-d_i).c_i\log\tau_1-2(x-d_i).f_i\frac{\log\tau_1}{\tau_1}+c_i^2(\log \tau_1)^2\big]^{1/2}.
\end{align}
We can expand the squareroot :
\begin{align}
\tau_2
&=\tau_0+(x-d_i).c_i \frac{\log \tau_0}{X}\ -\ c_i^2 \frac{(\log \tau_0)^2}{2X}+ (x-d_i).f_i\ \frac{ \log \tau_0}{X\tau^0} + (x.c_i-d_i.c_i)^2\frac{  (\log \tau_0)^2}{2X^3}\nn\\
&+  (x.c_i-d_i.c_i)^2\frac{ \log \tau_0\ }{\tau_0\ X^2}. \label{tau22}
\end{align}
We have used \eqref{tau11} for $\tau_1$ to derive above expression. And as before  $X=[\ (V_i.x-V_i.d_i)^2 +(x-d_i)^2\ ]^{1/2}$.
Now, let us study above expression in the limit $r\rightarrow \infty$ with $u$ finite. We have :
\begin{align}
\tau_2|_{\mathcal{I}^+}&=-\frac{u+q.d_i}{q.V_i}- \ \frac{q.c_i}{q.V_i} \ \log u+ q.f_i\ \frac{ \log u}{u}- \frac{(q.c_i)^2}{q.V_i}\ \frac{ \log u}{u}+\mathcal{O}(\frac{1}{r}).\label{tau2}
\end{align}
The $\mathcal{O}(\frac{1}{r})$ term starts at $\mathcal{O}(u^2)$. This produces $\mathcal{O}(\frac{u^2}{r^3})$-term in $A_\mu$ (see \eqref{A23}). We see from \eqref{tau22} that there is a  $\mathcal{O}(\frac{(\log u)^2}{r})$ term, this contributes to the $\mathcal{O}(\frac{(\log u)^2}{r^3})$-term in $A_\mu$.  
We can expand \eqref{tau22} in large $r$ limit keeping $v$ finite to get :
\begin{align}
\tau_2|_{\mathcal{I}^-}&=-2r\ V_i.\bar{q}+ \  \frac{\bar{q}.c_i}{V_i.\bar{q}} \ \log r +\frac{(\bar{q}.c_i)^2}{(V_i.\bar{q})^3} \frac{(\log r)^2}{2r} -\frac{c_i^2}{ \bar{q}.V_i} \frac{(\log r)^2}{2r} +\mathcal{O}(\frac{\log r}{r}). \label{t2p}
\end{align}

\section{Effect of long range forces on asymptotic trajectories}\label{lrf}
Let us find the first order correction to equation of trajectory of particles in asymptotic regions. This calculation has been done in \cite{Sen Sahoo}, we reproduce it here. The equation of trajectory of $j^{th}$  outgoing particle is given by :
\begin{align}
m_j\frac{\p^2x_j^\mu}{\p \tau^2}=e_j\ F^{\mu\nu}(x_j(\tau))\ V_{j\nu} .\label{e}
\end{align}
We need to find the field experienced by $j$ due to all other particles. We have calculated the field strength in \eqref{F1}. We evaluate the field strength \eqref{F1} at the position of the particle i.e. $x=x_j(\tau)$ :
\begin{align}
F_{\mu\nu}(x_j(\tau))
&=\frac{\Theta(\tau)}{4\pi \tau^2}\sum_{\substack{i=m+1,\\ i\neq j}} ^ne_i \frac{ (V_{i\mu}V_{j\nu}-V_{j\mu}V_{i\nu}) }{[(V_i.V_j)^2-1]^{3/2}} +\mathcal{O}(\frac{1}{\tau^3}) .\label{F}
\end{align}
Here, we have not included any incoming particles as they cannot affect the outgoing particles. So, the trajectory of the $j^{th}$ particle is given by following equation :
\begin{align}
m_j\frac{\p^2x_j^\mu}{\p \tau^2}=-\frac{e_j\Theta(\tau)}{4\pi \tau^2}\sum_{\substack{i=m+1,\\ i\neq j}} ^ne_i  \frac{  (V_{i\mu}+V_{j\mu}V_i.V_j)}{[(V_i.V_j)^2-1]^{3/2}} .\label{eot}
\end{align}
Here, we drop the $\mathcal{O}(\frac{1}{\tau^3})$ correction which is justified as we are working in large $\tau$ regime. But it is important to note it. Thus, asymptotic trajectories of the particles are corrected to :
\be x^\mu_i= V_i^\mu \ \tau +c^\mu_i \log \tau + d_i. \ee
where as given in \cite{Sen Sahoo}, for outgoing particles :
\begin{align}
c_i^\mu&=\frac{1}{4\pi} \sum_{\substack{j=m+1,\\ j\neq i}} ^n Q_iQ_j \frac{ p_i.p_j\ m_j^2 p^\mu_i+m_i^2m_j^2p_j^\mu}{[(p_i.p_j)^2-m_i^2m_j^2]^{3/2}}.\label{ca}
\end{align}
Above expression carries an extra minus sign compared to \cite{Sen Sahoo} because of difference in convention of $\eta_i$.
For $i^{th}$ incoming particle, $j$ runs over the incoming particles :
\begin{align}
c_i^\mu&=\frac{1}{4\pi} \sum_{\substack{j=1,\\ j\neq i}} ^m Q_iQ_j \frac{ p_i.p_j\ m_j^2 p^\mu_i+m_i^2m_j^2p_j^\mu}{[(p_i.p_j)^2-m_i^2m_j^2]^{3/2}}.
\end{align}

\subsection{Subleading correction to the equation  of trajectory}

Next we find the subleading correction to the equation of trajectory \eqref{eot}. Using \eqref{A01}, we calculate the field strength and evaluate it at the position of $j^{th}$ particle. We get following modes in the field strength 
\be  F_{\mu\nu}(x_j(\tau)) \sim e\ \sum_{m=2}^\infty\frac{1}{\tau^m}+e^3\ \sum_{m=3}^\infty \frac{\log\tau}{\tau^m}.\label{F22} \ee
Hence, the corrections to \eqref{eot} are of the form : 
\begin{align}
m_j\frac{\p^2x_j^\mu}{\p \tau^2}\sim \frac{e^2}{\tau^2} +e^4\frac{\log\tau}{\tau^3} +\frac{e^2}{\tau^3} +\cdots.\label{e1}
\end{align}
Thus, $\frac{\log\tau}{\tau^3}$ represents the subleading correction (at large $\tau$) to the trajectory. This $\frac{\log\tau}{\tau^3}$ term arises from the $\frac{\log u}{r^3}$-like modes in $F_{\mu\nu}$. Using \eqref{A01}, we get :
\begin{align}
F_{\mu\sigma}(x)|_{u\rightarrow\infty}
&=\frac{1}{4\pi }\sum_{i=m+1}^n   \frac{ e_i}{X^3}    \Big[-( V_{i\sigma}x_\mu-x_\sigma  V_{i\mu})\big[\frac{ 3x.c_i }{X^2}\log  u+1\big] + ( V_{i\sigma}c_{i\mu}-c_{i\sigma}  V_{i\mu})\ \log u\  \Big].\label{F2}
\end{align}
In above expression, we have retained only the relevant terms of $F_{\mu\sigma}(x)$. These modes in $F_{\mu\sigma}(x)$ are part of the radiation emitted by particles as the result of the particles undergoing acceleration under long range forces. Now as second order effect, a particle $j$ will accelerate as a result of the radiation emitted. To find the effect of field generated by other particles at the poistion of $j^{ th}$ particle  we substitute :
\be x^\mu=x^\mu_j(\tau)= V_j^\mu \tau +c^\mu_j \log \tau +d_j^\mu.\nn \ee
Substituting in \eqref{F2}, we get :
\begin{align}
&F_{\sigma\mu}(x_j)\nn\\
&=\sum_{\substack{i=m+1,\\ i\neq j}}^n\frac{e_i}{4\pi \tau^3}  \frac{\log \tau}{[(V_i.V_j)^2-1]^{3/2}}    \Big[3(V_j.c_i-V_i.c_j\ V_i.V_j)\frac{( V_{i\sigma}V_{j\mu}-V_{j\sigma}  V_{i\mu})}{[(V_i.V_j)^2-1]}+ ( V_{i\sigma}c_{j\mu}-c_{j\sigma}  V_{i\mu}) - ( V_{i\sigma}c_{i\mu}-c_{i\sigma}  V_{i\mu})  \Big].
\end{align}
There are terms depending on $c_j$ i.e. acceleration of $j^{ th}$ particle itself. 

Finally we have :

\begin{align}
m_j\frac{\p^2x_j^\mu}{\p \tau^2}
& =-\sum_{\substack{i=m+1,\\ i\neq j}}^n \frac{ e_ie_j\ (\log \tau)}{\tau^3[(V_i.V_j)^2-1]^{3/2}}\Big[ 3(V_j.c_i-V_i.c_j\ V_i.V_j)\  \frac{( V_i^\mu+V^\mu_{j}V_i.V_j  )}{[(V_i.V_j)^2-1]}+  V_{i}.V_jc_{j}^{\mu}- ( V_{i}.V_jc^\mu_{i}-V_j.c_{i}  V^\mu_{i}) \Big].
\end{align}
Thus, asymptotic trajectories of the particles are corrected to :
\be x^\mu_i= V^\mu_i \tau +c^\mu_i \log \tau + d_i+\eta_if_{i\sigma}\frac{\log\tau}{\tau}, \label{x2} \ee
where
\begin{align} f^\mu_i = -\sum_{\substack{j=m+1,\\ i\neq j}}^n m_im_j^2\frac{Q_iQ_j}{2} \Big[ 3(m^2_jp_i.c_j-p_j.c_i\ p_i.p_j)\  \frac{( p_i.p_j\ p_i^\mu+ m_i^2 p^\mu_{j})}{[(p_i.p_j)^2-m_i^2m_j^2]^{5/2}}\ + \frac{[ p_i.p_j\ c^\mu_{i}-( p_i.p_j\ c^\mu_{j}-p_i.c_j\  p^\mu_{j})]}{[(p_i.p_j)^2-m_i^2m_j^2]^{3/2}} \Big].  \label{f}\end{align}
For incoming particles, $f^\mu_i$ is negative of above expression. So, we have used $\eta_if_i^\mu$.\\

\section{Effect of internal structure and non minimal couplings}\label{int}
As we noted at beginning of section 2 the internal structure of the scattering objects does not affect the asymptotic modes that we study. Let us illustrate this point by studying the effect of dipole moment of the scattered objects on asymptotic field. It is obvious that $T$ has to be much larger than the order of the sizes of these objects. The bodies might not have uniformly distributed charges, the respective charge distribution is goven by a function $\rho_i(x')$. Any localised charge density function has a multipole expansion that goes as follows \cite{namias,kocher}
$$\rho_i(x') = e_i\ \delta^4(x'-x_i) + e_i\ a_i^\mu \p_\mu\delta^4(x'-x_i)+...\ .$$
The first term denotes the monopole contribution, the second term denotes the dipole contribution here we restrict to the dipole moment; subsequent calculation will show us that the higher order moments can be ignored. The information of the body is contained in coefficients like $a_i^\mu$.

 To the leading order in $e$, the trajectories of the centre of mass of the objects in the asymptotic regions is given by
$$ x^\mu_i= [V_i^\mu \tau + d_i]\Theta(-T-\tau).$$
$\tau$ is an affine parameter. Similarly, an outgoing particle has the trajectory :
$$ x^\mu_j= [V_j^\mu \tau + d_j]\Theta(\tau-T).$$
 The asymptotic current is given by :
\begin{align}
j^{\text{asym}}_\sigma(x') &= \int_T^\infty d\tau\sum_{i=n'+1}^n e_i V_{i\sigma}\ [\  \delta^4(x'-x_i)+\  a_i^\mu\p'_\mu\delta^4(x'-x_i)]\ \nn\\
&+\int_{-\infty}^{-T} d\tau\sum^{n'}_{i= 1} e_i V_{i\sigma}\  [\  \delta^4(x'-x_i)+\  a_i^\mu\p'_\mu\delta^4(x'-x_i)].\label{d}
\end{align}
Let us first consider the contribution from outgoing trajectories. Using the retarded propagator, we get :
\begin{align}
A^{\text{asym}}_\sigma(x)
&=\frac{1}{2\pi }\int d^4x'\int_{T}^{\infty} d\tau\  \sum_{i=n'+1}^n e_i\ V_{i\sigma}\ [\  \delta^4(x'-x_i)+\  a_i^\mu\p'_\mu\delta^4(x'-x_i)]\  \delta([x-x']^2)\  \Theta(t-t')
\end{align}
The contribution from the dipole term is :
\begin{align}
A_{D\sigma}&=-\frac{1}{2\pi }\int_{T}^{\infty} d\tau\  \sum_{i=n'+1}^n e_i\ V_{i\sigma}\   a_i^\mu\  \p'_\mu\delta([x-x']^2)\  \Theta(t-t') |_{x'=x_i}\nn\\
&=-\frac{1}{2\pi }\ a_i^\mu\p_{i\mu}\int_{T}^{\infty} d\tau\  \sum_{i=n'+1}^n e_i\ V_{i\sigma}\   \frac{\delta(\tau-\tau_0)}{|2\tau+2V_i.(x-d_i)|}\nn\\
&=-\frac{1}{2\pi }\ a_i^\mu\p_{i\mu}  \sum_{i=n'+1}^n e_i\ V_{i\sigma}\   \frac{1}{|2\tau_0+2V_i.(x-d_i)|},
\end{align}
using $\tau_0=-V_i.(x-d_i)-\big[\ (V_i.x-V_i.d_i)^2 +(x-d_i)^2\ \big]^{1/2}$, we see that 
$$A^D_{\sigma}\sim \mathcal{O}(\frac{1}{r^2}).$$
From above calculation it is clear that $m^{th}$ order moment starts contributing at $\mathcal{O}(\frac{1}{r^m})$. This is discussed in elementary books on electrodynamics.

Let us include the effect of long range force on asymptotic trajectory. The equation of trajectory of centre of mass of $j^{th}$ body is given by \eqref{e}
\begin{align}
m_j\frac{\p^2x_j^\mu}{\p \tau^2}=e_j\ F^{\mu\nu}(x_j(\tau))\ V_{j\nu} \ .
\end{align}
We have
\begin{align}
F_{\mu\nu}(x_j(\tau))
&=\sum_{n=2}^\infty\frac{c_n}{\tau^n},
\end{align}
such that $m^{th}$ order moment starts contributing at $\mathcal{O}(\frac{1}{\tau^{m+1}})$. We can repeat the analysis of \ref{lrf} to show that the centre of mass of the objects follow trajectories of the form 
\be x^\mu_j(\tau)= V_j^\mu \tau +c^\mu_j \log \tau +d_j^\mu+\mathcal{O}(\frac{1}{\tau}).\nn \ee
The higher order moments contribute to above equation at $\mathcal{O}(\frac{1}{\tau})$ and do not modify the expression of $c_j^\mu$ given in \eqref{x1}. 
 Including the effect of long range force, the asymptotic current is given by :
\begin{align}
j^{\text{asym}}_\sigma(x') &= \int_T^\infty d\tau\sum_{i=n'+1}^n e_i [V_{i\sigma}+\frac{c_{i\sigma}}{\tau}]\ [\  \delta^4(x'-x_i)+\  a_i^\mu\p'_\mu\delta^4(x'-x_i)]\ \nn\\
&+\int_{-\infty}^{-T} d\tau\sum^{n'}_{i= 1} e_i [V_{i\sigma}+\frac{c_{i\sigma}}{\tau}]\  [\  \delta^4(x'-x_i)+\  a_i^\mu\p'_\mu\delta^4(x'-x_i)].
\end{align}
We can easily calculate the radiative field produced by above current. The contribution of the dipole term has following structure :
$$A^D_{\sigma}\sim \frac{1}{r^2}\Big[e\ u^0+e^3\ \frac{1}{u}+...\Big] + \frac{1}{r^3}\Big[e\ u+e^3\ \log u+...\Big] .$$
Calculating the field strength (which involves derivative of above expression), we get
\begin{align}
F^D_{\mu\nu}
&\sim \frac{1}{r^3}\Big[e\ u^0+e^3\ \frac{1}{u}+...\Big] + \frac{1}{r^3}\Big[e\ u+e^3\ \log u+...\Big] . \label{FD1}
\end{align}
Hence the calculation in section 3 is not modified by the dipole term i.e. $Q_1$ charge in \eqref{1loop} is unaffected.\\

Next we need to check if the subleading tail term given in \eqref{Alogu} and the $Q_2$ charge given in \eqref{2loop} are modified. We have already shown that $c_j^\mu$ term in \eqref{x22} is unaffected by dipole moments, if we show that $f_j^\mu$ term in \eqref{x22} is also unaffected then it follows that the the subleading tail term and the $Q_2$ charge are not modified.

Let us the $\mathcal{O}(e^4)$ corrections to the equation of trajectory. We evaluate the field strength in \eqref{FD1} at the position of $j^{th}$ body to get
\begin{align}
F^D_{\mu\nu}(x_j(\tau))
&\sim e\frac{1}{\tau^3}+e^3\frac{\log\tau}{\tau^4}+...\ . \label{FD}
\end{align}

Hence, analogous to \eqref{e1}, we get following equation for asymptotic trajectory of centre of mass of the $j^{th}$ body :
\begin{align}
m_j\frac{\p^2x_j^\mu}{\p \tau^2}\sim \frac{e^2}{\tau^2} +e^4\frac{\log\tau}{\tau^3} +\frac{e^2}{\tau^3} ++e^4\frac{\log\tau}{\tau^4}+\cdots.
\end{align}
From \eqref{FD}, we see that the dipole moment cannot contribute to the first two terms on the RHS. Hence $c_j^\mu, f_j^\mu$ terms in \eqref{x22} are unaffected by dipole moments. Using similar argument it can proved that higher order moments too cannot affect these terms. 

Thus the leading $\frac{1}{u}$ tail (and the $Q_1$ charge) as well as the subleading $\frac{\log u}{u^2}$ tail (and the $Q_2$ charge) are unaffected by higher order moments of the scattering particles.\\\\ 

\textbf{Effect of non minimal couplings}\\
The effect of a general non-minimal coupling of $U(1)$ gauge field to equation of trajectory point particle takes following form :
\begin{align}
m_j\frac{\p^2x_j^\mu}{\p \tau^2}=e_j\ F^{\mu\nu}(x_j(\tau))\ V_{j\nu} +\  C_{\lambda\nu}\ \p^\mu F^{\lambda\nu}(x_j(\tau)) .
\end{align}
Here, $C_{\lambda\nu}$ is an arbitrary tensor depending on $x_j(\tau)$ or it could also contain more factors of $F_{\mu\nu}$. Above correction originates from following correction to the current.
\begin{align}
j^{\text{asym}}_\sigma(x') &= \int_T^\infty d\tau\sum_{i=n'+1}^n [e_i V_{i\sigma}\  \delta^4(x'-x_i)+\  C_{\sigma}^{\ \mu}\p'_\mu\delta^4(x'-x_i)]\ \nn\\
&+\int_{-\infty}^{-T} d\tau\sum^{n'}_{i= 1} e_i V_{i\sigma}\  [e_i V_{i\sigma}\  \delta^4(x'-x_i)+\  C_{\sigma}^{\ \mu}\p'_\mu\delta^4(x'-x_i)].
\end{align}
We can compare the non-minimal term with the dipole term in \eqref{d}. Thus repeating similar calculation, we can show that the non-minimal term will not affect $f_{i\sigma}$ and $c_{i\sigma}$ in \eqref{x2}. Thus the leading $\frac{1}{u}$ tail (and the $Q_1$ charge) as well as the subleading $\frac{\log u}{u^2}$ tail (and the $Q_2$ charge) are unaffected by non-minimal couplings.\\\\

\textbf{Effect of internal spin}\\
Let us discuss the equation of trajectory of a point particle that carries an internal spin. It takes following form \cite{spin}
\begin{align}
m_j\frac{\p^2x_j^\mu}{\p \tau^2}=e_j\ F^{\mu\nu}(x_j(\tau))\ V_{j\nu} +\  S_{j\lambda\nu}\ \p^\mu F^{\lambda\nu}(x_j(\tau)) .
\end{align}
Here, $S_{j\lambda\nu}$ is the internal spin tensor of the $j^{th}$ body. Thus we see that the coupling of internal spin of the particle is a special type of non-minimal coupling. This shows that the internal spins of the scattering objects do not affect our analysis of  thesubleading tail term and the expression of $Q_1,Q_2$ charges.

\section{Maxwell'equations at future null infinity}
The Maxwell's equations $\nabla^{\nu}F_{\sigma\nu}\ =\ j_{\sigma}$ can be expanded order by order around $\mathcal{I}^+$. We have already calculated the leading-$r$ terms in the field strength in \eqref{F1} :
\begin{align}
F_{rA} & =\frac{F^{[1/r^2]}_{rA}(u,\hat{x})}{r^2}+...\ , & F_{ru}  =\frac{F^{[1/r^2]}_{ru}(u,\hat{x})}{r^2}+...\ , && F_{uA}  =F^{[r^0]}_{uA}(u,\hat{x})+...\ , && F_{AB}  =F^{[r^0]}_{AB}(u,\hat{x})+...\ .\nn
\end{align} 
We substitute above fall-offs for the field strength components in Maxwell's equations, using Bianchi identities we get following equations :
\begin{align}
\partial_uF^{[1/r^2]}_{ru}\ +\  \partial_uD^BA_B^0\ =\ 0,\nn\\
\partial_uF^{[1/r^2]}_{rA}\ -\ \frac{1}{2}\partial_AF^{[1/r^2]}_{ru}\ +\frac{1}{2} D^BF_{AB}^{[r^0]}\ =\ 0.\label{maxwell}
\end{align}
Here, we have used the fact that massive currents decay very fast at $\mathcal{I}^+$ and the equations become homogenous. 
From \eqref{maxwell}, we have, \be \partial^2_uF^{[1/r^2]}_{rA}\ +\ \frac{1}{2}\partial_u\p_AD^BA_B^0\ +\frac{1}{2} \p_uD^BF_{AB}^{[r^0]}\ =\ 0. \label{m1}\ee
Here, $A_A^0$ denotes following behaviour in $A_A(x)$ : $A_A(x)\sim A_A^0(u,\hat{x}) + \mathcal{O}(\frac{1}{r})$. Above equation can be used to relate $1/u$-term in $A^0_A$ to the $\frac{\log u}{r^2}$ term in $F_{rA}$. In particular, the $z$ component of above equation gives us :
\begin{align}
F^{[\log u/r^2]}_{rz} =-D_z^2 A^{[1/u]}_{\bar{z}} .\label{101}
\end{align}
Above equation is the one of our interest as it relates the charge on the left side while the right side involves $A^{[1/u]}_{\bar{z}}$  that is proportional to the 1-loop soft factor.

Next we repeat above analysis for next order in $\frac{1}{r}$. Maxwell's equations lead to :
\begin{align}
&\partial_uF^{[1/r^3]}_{ru}\ +\ D^BF_{uB}^{[1/r]}\ =\ 0,\nn\\
&\partial_u{F}^{[1/r^3]}_{rA}-2F^{[1/r^2]}_{uA}+2F^{[1/r^2]}_{rA} + D^BF_{AB}^{[1/r]}=0\nn
\end{align}
We use Bianchi identities to eliminate all $F_{uA}$'s and $F_{AB}^{[1/r]}$. After some manipulation we arrive at following equation :
\begin{align}
&2\partial^3_u{F}^{[1/r^3]}_{rA} +\p_AD^2\ \p_uF_{ru}^{[1/r^2]}+\big[2\delta^B_A -\p_A D^B+\delta^B_AD^2-D^B\p_A\big]\p^2_uF^{[1/r^2]}_{rB}=0.
\label{maxwell2}
\end{align}
We substitute for $\p_uF_{ru}^{[1/r^2]}$ from \eqref{maxwell} and for $\p^2_uF^{[1/r^2]}_{rB}$ from \eqref{m1} in \eqref{maxwell2}, we get :
\be 2\p^3_uF^{[1/r^3]}_{rB}\ =\ \p_u\ [\mathcal{D}^B]\ A_B^0.\label{201}\ee
$\mathcal{D}^B$ is a function of sphere derivatives that contains upto fourth-order derivatives. Its exact form is not important for us. The takeaway point from this equation is that $(\log u)^2$ term in $F^{[1/r^3]}_{rB}$ is fixed in terms of $\frac{\log u}{u^2}$ of $A_B^0$.


\end{document}